\documentclass[12pt]{iopart}
\eqnobysec

\usepackage{iopams,graphicx,amssymb}

\def\eps{\varepsilon}
\def\Dm{\widetilde{\cal D}_{\mu}}
\def\D{{\cal D}}

\def\E{{\cal E}}
\def\p{{\bf p}}

\def\n{{\bf n}}

\begin{document}

\title[Renormalization group in the infinite-dimensional turbulence]
{Renormalization group in the infinite-dimensional turbulence:
Third-order results}

\author{L~Ts~Adzhemyan, N~V~Antonov, P~B~Gol'din, T~L~Kim and
M~V~Kompaniets}

\address{Department of Theoretical Physics, St.~Petersburg University,
Uljanovskaja 1, St.~Petersburg, Petrodvorez, 198504 Russia}

\ead{nikolai.antonov@pobox.spbu.ru}

\begin{abstract}
The field theoretic renormalization group is applied to the stochastic
Navier--Stokes equation with the stirring force correlator of the form
$k^{4-d-2\eps}$ in the $d$-dimensional space, in connection with the
problem of construction of the $1/d$ expansion for the fully developed
fluid turbulence beyond the scope of the standard $\eps$ expansion.
It is shown that in the large-$d$ limit the number of the Feynman diagrams
for the Green function (linear response function) decreases drastically,
and the technique of their analytical calculation is developed. The main
ingredients of the renormalization group approach -- the renormalization
constant, $\beta$ function and the ultraviolet correction exponent
$\omega$, are calculated to order $\eps^{3}$ (three-loop approximation).
The two-point velocity-velocity correlation function, the Kolmogorov
constant $C_{K}$ in the spectrum of turbulent energy and the inertial-range
skewness factor ${\cal S}$ are calculated in the large-$d$ limit to
third order of the $\eps$ expansion.
Surprisingly enough, our results for $C_{K}$ are in a reasonable
agreement with the existing experimental estimates.
\end{abstract}


\maketitle

\section{Introduction} \label{sec:Intro}

One of the most interesting open problems inherited by the modern
theoretical physics from the twentieth century is that of
description of fully developed hydrodynamic turbulence on the
basis of a microscopic model and within a consistent perturbation
scheme \cite{Legacy}.  A challenge which is still waiting to be
answered is the derivation of anomalous scaling behaviour of the
velocity correlation functions from first principles and
calculation of the corresponding anomalous exponents within a
regular perturbation theory, analogous to the $\eps$ or
$1/N$ expansions of the critical exponents in the theory of
second-order phase transitions.

The first difficulty is that the ordinary perturbation theory for the
stirred (stochastic) Navier--Stokes (NS) equation (that is, the expansion
in the nonlinearity) is in fact the expansion in the Reynolds number,
a parameter which tends to infinity for the fully developed turbulence.
Hence the necessity to rearrange (to sum up) the naive perturbation series.
A similar problem is well known in the theory of critical behaviour, where
it had been solved a long ago by means of the renormalization group
(RG) and the self-consistency (``bootstrap'') diagrammatic equations
\cite{Zinn,Book3}. The first way naturally leads to the famous $\eps$
expansion (where $\eps=4-d$
is the deviation of the spatial dimension $d$ from its upper critical
value $d=4$), the second leads to the alternative $1/N$ expansion (where
$N$ is the number of components of the corresponding order parameter).
So far, however, those methods have had relatively limited success when
applied to the problem of anomalous scaling in fluid turbulence.

An important difference is that the turbulence or, better to say,
the stochastic NS equation, has no upper critical dimension, and the
parameter $\eps$ in the standard RG
approach has completely different meaning. Namely, the correlation
function of the random stirring force that provides the energy supply
to the system is taken in the power-law form \cite{DeDom}
\begin{equation}
\langle ff \rangle \propto k^{4-d-2\eps},
\label{ff}
\end{equation}
where $k$ is the wave number; more precisely, see \cite{Book3,UFN,turbo}
and section~\ref{sec:QFT}.

The physical value $\eps=2$ corresponds to the energy pumping by
the largest scales, $\langle ff \rangle \propto \delta({\bf k})$, while
$d$ remains a free parameter and can be varied independently of $\eps$.
The common point with the models of critical behaviour is that, in the
both cases, the limit $\eps\to0$ corresponds to a logarithmic (exactly
renormalizable) theory and $\eps$ serves as the formal small expansion
parameter in the RG approach. For this reason, we use the same symbol
$\eps$ for the stochastic NS problem (where it is also sometimes denoted
as $\eps=y/2$).

The results of the RG analysis of this model are reliable and internally
consistent for asymptotically small $\eps$, while the possibility of their
extrapolation to the physical value $\eps=2$ and thus their relevance for the
real fluid turbulence is far from obvious. Of course, the physical value
of $\eps=4-d$ in the RG theory of critical phenomena is not small, either.
But there, no qualitative changeover in the behaviour of the system is
expected when $\eps$ increases from the region $\eps\ll 1$ to real values
$\eps\sim 1$, and the possibility of this extrapolation is usually not
disputed. The situation in the RG approach to the NS model (\ref{ff})
appears different. New physical effects are encountered as $\eps$ grows,
and they can be easily lost or misrepresented if the $\eps$ expansion
is too naively applied. One of them is related to the so-called
sweeping effects, the transport of small turbulent eddies as a whole by
large-scale ones. The sweeping leads to strong dependence of the velocity
correlation functions of the integral (external) turbulence scale $L$.
The other effect is the crossover from the Kolmogorov ``K41''scaling
to the anomalous (multi-)scaling -- singular dependence of the Galilean
invariant quantities (like e.g. equal-time structure functions) on $L$,
characterized by an infinite set of independent anomalous exponents.
These effects lead to infrared (IR) divergences in the diagrams of
perturbation theory which, formally speaking, manifest themselves as poles
at some finite values of $\eps$. Such poles and their physical interpretation
were discussed in a number of papers; see \cite{K}--\cite{APS}.

The aforementioned crossovers have no analog in the models of critical
behaviour, but such IR poles are present in their diagrams, too. Moreover,
they approach closer and closer to the origin $\eps=0$ when the order of the
perturbation theory (complexity of the corresponding diagrams) increases.
The correct resummation of these IR singularities is accomplished by the
short-distance operator-product expansion (SDE); it shows that the finite
limit $L\to\infty$ (where $L$ is some IR scale) in the correlation functions
exists and the singular-in-$L$ terms give only subleading corrections to the
scaling behaviour \cite{Zinn,Book3}. Thus the existence of IR singularities
at finite values of $\eps$ does not hinder the use of the RG technique and
the $\eps$ expansion for the description of critical behaviour.

The SDE technique is equally applied to the stochastic NS problem.
The distinguishing feature, specific to models of turbulence, is the
existence in the corresponding SDE of composite fields (``operators'')
 with {\it negative} scaling dimensions. Such operators were named
``dangerous'' in \cite{JETP} because their contributions to the OPE
diverge for $L\to\infty$. The summation of the most singular contributions
coming from the operators $v^n$, powers of the velocity field, was
performed in Ref. \cite{JETP}, the generalization to the case of a
time-dependent large-scale field is given in \cite{Kim2}. This gives
the adequate description of the sweeping effects within the RG formalism;
see also the general discussion in Refs. \cite{UFN,turbo}.

According to the SDE scenario, anomalous multiscaling in the
structure functions can be related to the existence of Galilean
invariant dangerous operators. This idea was successfully realized
for the model of a scalar impurity field passively advected by the
Gaussian velocity field with given correlation function
$\propto\delta(t-t')/k^{d+\eps}$, known as the Obukhov--Kraichnan
rapid-change model. In the zero-mode approach, which can be viewed
as a variant of the self-consistency equations, the anomalous
exponents were derived analytically to order $O(\eps)$ \cite{GK}
and $O(1/d)$ \cite{Falk1}; see also \cite{FGV} for a detailed
discussion. In the RG and SDE approach, the exponents are
identified with the scaling dimensions of composite operators
built of the scalar gradients (namely, powers of the dissipation
rate of the scalar field fluctuations, and their tensor analogs)
and calculated within the expansion in the exponent $\eps$ (also
denoted by $\xi$) to order $\eps^{3}$ (three-loop approximation of
the RG); see the original works \cite{RG1,RG3} and the review paper
\cite{JphysA} for generalizations and more references.

In the stochastic NS problem, however, all the critical dimensions
are strictly positive for small $\eps$ and can become negative only for
some finite values of $\eps$. Therefore, dangerous invariant operators
cannot be identified within the $\eps$ expansions and it is desirable to
construct an alternative perturbation scheme valid for finite $\eps$.
Attempts were made to modify the model by introducing $N$ replicas of the
velocity field and to construct an expansion in $1/N$ \cite{WM},
but such modifications were inconsistent with the Galilean symmetry
\cite{Gallina}; see also discussion \cite{Eyink,Daniela}
for the case of discrete shell models.

Thus we naturally return to an old idea of the expansion in $1/d$,
which has repeatedly been introduced in various contexts in turbulence
\cite{OK,FFR,YakhotD,HLF,Anton,vector} and looks very attractive
for a few reasons.

One can hope that in the limit $d\to\infty$ intermittency and anomalous
scaling disappear or acquire a simple ``calculable'' form and the
finite-dimensional turbulence can be studied within the expansion around
this ``solvable'' limit \cite{FFR}. However, in contrast to the
$O(N)$-symmetric model of the critical behaviour, where the limit
$N\to\infty$ is described in a closed form by the exactly solvable spherical
model \cite{sfera}, no drastic simplifications were found in \cite{FFR}
for the infinite-dimensional turbulence: all classes of diagrams for the
velocity correlation function (for nonstationary perturbation theory with
the Taylor expansion in time and for renormalized perturbation theory
with skeleton diagrams with dressed lines) survive in the large-$d$ limit,
and the incompressibility condition (and thus the nonlocal pressure effects)
remain important. Much later it was argued (on the basis of a certain
SDE-motivated ansatz for dissipative terms) that the K41 theory becomes
exact and the multiscaling indeed disappears for $d=\infty$ \cite{YakhotD},
as it also happens for the Obukhov--Kraichnan model \cite{OK}. What is more,
for the latter it was possible to find the $O(1/d)$ contribution to the
anomalous exponents \cite{Falk1}. Numerical simulation of the passive scalar
advection in the Obukhov--Kraichnan model for large $d$ (up to $d=30$)
was performed in \cite{Paolo}. However, the systematic expansion in $1/d$
has not been yet constructed for that model, let alone the
stochastic NS equation.

The key idea of the present paper is to combine the large-$d$
limit with the RG approach and the expansion in $\eps$.
It was noticed earlier in \cite{Anton,vector} that taking the limit
$d\to\infty$ leads to serious simplifications in the RG
calculations. In a very important paper \cite{Anton}, scaling
dimensions of all the powers of the local energy dissipation rate
were calculated for $d=\infty$ to first order in $\eps$. For a
finite $d$, such calculation becomes an extremely daunting task
due to the mixing of operators in the renormalization, and it was not
completely performed even for the second power of the dissipation rate;
see \cite{Eight}. The scaling dimensions calculated in
\cite{Anton} appear positive for $\eps<2$ and vanish at the
physical value $\eps=2$, in agreement with the arguments of Refs.
\cite{FFR,YakhotD} that the K41 scaling exactly holds for
$d=\infty$. It is then possible that the $O(\eps/d)$ correction to
the result of \cite{Anton} will make the dimensions negative and
the anomalous scaling will be established as a fact reliable
within the double expansion in $\eps$ and $1/d$ (we recall that the
powers of the dissipation rate are the operators ``responsible'' for the
anomalous scaling in the Obukhov--Kraichnan model \cite{RG1,RG3,JphysA}).
For the vector analog of the rapid-change model, where the calculation
of the anomalous exponents also becomes rather involved already in the
$O(\eps)$ order due to the mixing of composite fields, additional
expanding in $1/d$ also leads to serious simplification \cite{vector}.

In this paper we systematically investigate the infinite-$d$ limit
within the RG framework. Although we were not able to exactly solve the
problem for $d=\infty$ nor to construct an analog of the ``spherical
model,'' we obtained a number of new results which we consider interesting
and which, as we believe, can give some hints for the further study of the
large-$d$ behaviour.

In section~\ref{sec:QFT} we briefly recall the field theoretic formulation,
renormalization and the main findings of the RG approach for the stirred NS
equation. A more detailed specification of the limit  $d\to\infty$ is given.
In section~\ref{sec:Grin}
we consider the linear response (Green) function and show that, for
$d\to\infty$, lots of diagrams vanish and the remaining ones simplify
drastically. This allows us to analytically calculate in the three-loop
approximation the key ingredients of the RG approach: the renormalization
constant, $\beta$ function, coordinate of the fixed point and the
ultraviolet (UV) correction exponent $\omega$ (for finite $d$, only
two-loop results for these quantities are known, and they involve
{\it numerical} calculations of the corresponding diagrams \cite{APS}).
Some speculations are also made about the hypothetical expressions for
these quantities beyond the scope of the $\eps$ expansion.

The two-point velocity-velocity correlation function is analyzed in
section~\ref{sec:pair}. Here, no remarkable simplifications in the diagrams
occur in the limit $d\to\infty$, in agreement with the observations made
earlier in Ref.~\cite{FFR}. However, we propose some trick which allows us
to perform the analytical calculation of the pair correlator in the two-loop
approximation (which, in orders of the coupling constant, corresponds to
the three-loop approximation for the Green function). Although the results
can be interpreted {\it a posteriori\,} in terms of the steepest-descent
method, the practical use of the latter beyond the one-loop approximation
is hardly possible due to complexity of the corresponding integrals.

These results are exploited in section~\ref{sec:CK}. There, we calculate
to the third order in $\eps$ the Kolmogorov constant $C_{K}$ in the spectrum
of turbulent energy and the inertial-range skewness factor ${\cal S}$.
An important point here is not only the inclusion of the third-order
correction, but also the derivation of $C_{K}$ through a universal
(in the sense of the theory of critical behaviour) quantity. This
approach was proposed earlier in \cite{APS}, where $C_{K}$ was calculated
(for $d=3$) to the second order in $\eps$. It allows one to obviate
the main shortcoming of earlier calculations of $C_{K}$:
the intrinsic ambiguities in the corresponding $\eps$ expansions
(see e.g. \cite{Lam} and the discussion in \cite{turbo}). The results are
briefly summarized in the Conclusion.

\section{Stochastic Navier--Stokes equation, choice of the random force
and the field theoretic formulation} \label{sec:QFT}

As the microscopic dynamical model of the fully developed, homogeneous,
isotropic turbulence of an incompressible viscous fluid one usually takes
the stochastic NS equation with a random stirring force
\begin{equation}
\nabla _t v_i=\nu _0\partial^{2} v _i-\partial _i {\cal P}+f_i,
\qquad \nabla _t = \partial _t + v_i \partial _i,
\label{1.1}
\end{equation}
where $v_i$ is the transverse (divergence-free, due to the incompressibility
condition $\partial_i v_i=0$) velocity field, ${\cal P}$ and $f_i$ are the
pressure and the transverse random force per unit mass (all these quantities
depend on $x= \{t,{\bf x} \}$), $\nu_0$ is the kinematic
viscosity coefficient, $\partial^{2}$ is the Laplacian and $\nabla_t$ is
the Lagrangian derivative. The problem (\ref{1.1}) is studied on the entire
$t$ axis and
is augmented by the retardation condition and the condition that
$v_i$ vanishes for $t\to-\infty$.

We assume for $f$ a Gaussian distribution with zero mean and correlation
function
\begin{equation}
\big\langle f_i(x)f_j(x')\big\rangle = \frac{\delta (t-t')}{(2\pi)^{d}}\,
\int d{\bf k}\, P_{ij}({\bf k})\, d_f(k)\, \exp \big[{\rm i}{\bf k}
\left({\bf x}-{\bf x}'\right)\big] ,
\label{1.2}
\end{equation}
where $P_{ij}({\bf k}) =\delta _{ij}  - k_i k_j / k^2$ is the transverse
projector, $d_f(k)$ is some function of the wave number $k=|{\bf k}|$
and model parameters, and $d$ is the dimension of the ${\bf x}$ space.
The time decorrelation of the random force guarantees Galilean invariance
of the stochastic problem (\ref{1.1}), (\ref{1.2}).

Let us specify the form of the function $d_f(k)$ in the correlator
(\ref{1.2}) used in the RG theory of turbulence. Physically, the random force
models the injection of energy into the system due to interaction with the
large-scale motions. Idealized injection by infinitely large vortices
corresponds to
\begin{equation}
d_f(k)= 2 (2\pi)^{d}\, \E\, \delta({\bf k}) / (d-1),
\label{2.75}
\end{equation}
where $\E$ is the average power of the injection (equal to the average
dissipation rate) and the amplitude factor comes from the exact relation
\begin{equation}
\E = \frac{(d-1)}{2 (2\pi)^{d}} \,\int d {\bf k} \, d_f(k).
\label{1.3}
\end{equation}
On the other hand, for the use of the standard RG technique it is important
that the function $d_{f}(k)$ have a power-law behaviour at large $k$. This
condition is satisfied if $d_{f}(k)$ is chosen in the form
\begin{equation}
d_f(k)=D_0\,k^{4-d-2\varepsilon},
\label{1.10}
\end{equation}
where $D_0>0$ is the amplitude factor and $\varepsilon>0$ is the exponent
with the physical value $\varepsilon=2$. This can be explained with the aid
of the well-known power-law representation of the $d$-dimensional $\delta$
function
\begin{eqnarray}
\delta({\bf k}) &=& \lim_{\eps\to 2}\, \frac{1}{(2\pi)^{d}} \int
d{\bf x} \, (\Lambda x)^{2\eps-4} \, \exp [{\rm i} ({\bf kx})] =
\nonumber \\ &=&
S_{d}^{-1} k^{-d} \lim_{\eps\to 2} \left[ (4-2\eps)
(k/\Lambda)^{4-2\eps} \right],
\label{2.76}
\end{eqnarray}
with some UV momentum scale $\Lambda$. Here and below we denote
\begin{eqnarray}
S_d = 2\pi^{d/2}/\Gamma (d/2), \qquad \bar S_d = S_d / (2\pi)^{d},
\label{surface}
\end{eqnarray}
where $S_d$ is the surface area of the unit sphere in $d$-dimensional space
and $\Gamma(\dots)$ is the Euler Gamma function.
For $\eps\to2$, the function (\ref{1.10}) turns to the
ideal injection (\ref{2.75}) if the amplitude $D_{0}$ is related to $\E$ as
\begin{eqnarray}
D_{0} \to \frac{4(2-\eps)\,\Lambda^{2\eps-4}}
{\overline S_{d}(d-1)}\,\, \E \qquad {\rm for} \ \eps\to2.
\label{2.74}
\end{eqnarray}
As already mentioned, in the RG approach to the problem (\ref{1.1}),
(\ref{1.2}), (\ref{1.10}) the exponent $\varepsilon$ plays the part
analogous to that played by $4-d$ in Wilson's theory of critical phenomena,
while $d$ remains a free parameter. A more realistic model is
\begin{equation}
d_f(k)=D_0\,k^{4-d-2\varepsilon}\, h(m/k), \qquad  h(0)=1,
\label{1.9}
\end{equation}
where $m=1/L$ is the reciprocal of the integral turbulence scale $L$ and
$h(m/k)$ is some well-behaved function that provides the IR regularization.
Its specific form is unessential; we shall always use the sharp cutoff
\begin{equation}
h(m/k)=\Theta(k-m),
\label{1.99}
\end{equation}
which is the most convenient choice from the calculational viewpoints;
$\Theta(\dots)$ being the Heaviside step function.

According to the general theorem \cite{MSR}, stochastic problem (\ref{1.1}),
(\ref{1.2}) is equivalent to the field theoretic model of the doubled set of
fields $\Phi= \{{\bf v}', {\bf v} \}$ with action functional
\begin{equation}
S(\Phi )= v 'D_{v}v'/2+ v'\left[-\nabla_t +\nu _0\partial^{2} \right]v,
\label{action}
\end{equation}
where $D_{v}$ is the correlation function (\ref{1.2}) of the random force
$f_{i}$ with the function $d_f(k)$ from (\ref{1.9}) and all the
required integrations over $x=\{t,{\bf x}\}$ and summations over the vector
indices are understood, for example,
\[ v'(v\partial)v = \int dt \int d{\bf x} \ v_{i}'
(v_{j}\partial_{j}) v_{i}. \]
The auxiliary vector field is also
transverse, $\partial_{i}v_{i}'=0$, which allows to omit the
pressure term on the right-hand side of Eq. (\ref{action}), as
becomes evident after the integration by parts:
$$ \int dt \int d{\bf x} \ v_{i}'\partial_{i} {\cal P} = - \int dt
\int d{\bf x} \ {\cal P} (\partial_{i}v_{i}') =0 . $$
Of course, this does not mean that the pressure contribution can simply be
neglected: the field $v'$ acts as the transverse projector and selects the
transverse part of the expressions to which it is contracted
in~(\ref{action}).

Formulation (\ref{action}) means that statistical averages of random
quantities in the original stochastic problem (\ref{1.1}), (\ref{1.2})
can be represented as functional averages with the weight $\exp S(\Phi)$,
and the generating functionals of total [$G(A)$] and connected [$W(A)$]
correlation functions of the problem are represented by the functional
integral
\begin{equation}
G(A)=\exp  W(A)=\int {\cal D}\Phi \exp [S(\Phi )+A\Phi ]
\label{14}
\end{equation}
with arbitrary sources $A = \{A^{v'},A^{v} \}$ in the linear form
$A\Phi = \sum _{\Phi}\int dx \,A^{\Phi}(x)\Phi (x)$.

The model (\ref{action}) corresponds to a standard Feynman diagrammatic
technique; the bare propagators (lines in the diagrams) in the
time--momentum ($t$--${\bf k}$) representation have the forms
\begin{eqnarray}
\bigl\langle v_{i}(t)  v_{j}'(t') \bigr\rangle _0 = \Theta(t-t')
\exp \left\{ - \nu_0 k^{2} (t-t') \right\} \, P_{ij}({\bf k})
=\raisebox{-0.05cm}{\includegraphics[height=0.3cm,width=1.5cm]{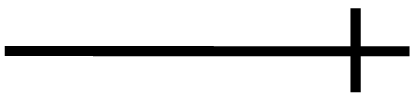}},
\nonumber \\
\bigl\langle v_{i}(t) v_{j}(t') \bigr\rangle _0 =
\frac{d_{f}(k)}{2\nu_0 k^{2}}\,
\exp \left\{ - \nu_0 k^{2} |t-t'| \right\} \, P_{ij}({\bf k})
=\raisebox{0.05cm}{\includegraphics[height=0.03cm,width=1.5cm]{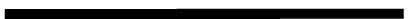}},
\nonumber \\
\bigl\langle v_{i}'(t) v_{j}'(t') \bigr\rangle _0 = 0,
\label{linesV}
\end{eqnarray}
with $d_{f}(k)$ from (\ref{1.9}) and the Heaviside step function
$\Theta(\dots)$. The interaction in (\ref{action}) corresponds to the triple
vertex $-v'(v\partial)v= v'_{i}V^{S}_{ijs}v_{j}v_{s}/2$ with vertex factor
\begin{equation}
V^{S}_{ijs} = {\rm i} (k_{j}\delta_{is}+k_{s}\delta_{ij})
=\raisebox{-0.40cm}{\includegraphics[height=1.0cm,width=1.0cm]{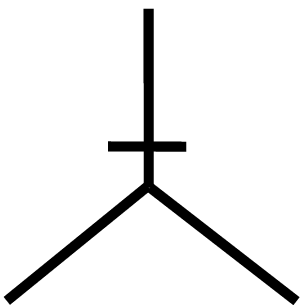}},
\label{vertexV}
\end{equation}
where ${\bf k}$ is the momentum argument of the field $v'$.

It is convenient to introduce the new parameter (``coupling constant'')
by the relation
\begin{equation}
g_{0} = D_{0}/\nu_0^3,
\label{Prad}
\end{equation}
so that $g_{0} \propto \Lambda^{2\eps}$ with the UV momentum scale
$\Lambda$ coming from (\ref{2.76}). Thus the model (\ref{action})
becomes logarithmic (the coupling constant becomes dimensionless)
at $\eps=0$, and the UV divergences manifest themselves as the
poles in $\eps$ in the correlation functions of the fields $\Phi =
\{v,v'\}$. Dimensional analysis (power counting) shows that
superficial UV divergences, whose removal requires counterterms,
can be present only in the 1-irreducible correlation functions
$\langle v'v\rangle_{\rm 1-ir}$ and the corresponding counterterms reduce
to the forms $v'\partial^{2}v$, $v'\partial_{t}v$ and $v'(v\partial)
v$. In fact, in the model (\ref{action}) there are fewer possible
counterterms than allowed by the naive power counting. The symbol
$\partial$ at the vertex in (\ref{action}) can be moved onto the
field $v'$ using the integration by parts, which means that the
counterterms to the 1-irreducible functions must contain at least
one spatial derivative per each field $v'$. This excludes the
structure $v'\partial_{t}v$. The Galilean symmetry requires that
the counterterms $v'\partial_{t}v$ and $v'(v\partial) v$ must form
the invariant combination $v'\nabla_{t}v$, which excludes the
structure $v'(v\partial) v$. In the special case $d=2$ a new UV
divergence appears in the 1-irreducible function $\langle
v'v'\rangle_{\rm 1-ir}$. Since we are interested in the large-$d$
limit here, we can simply ignore that divergence.

Then the inclusion of the only remaining counterterm $v'\partial^{2}v$
in the action functional (\ref{action}) is reproduced by the multiplicative
renormalization of the parameters $\nu_0$ and $g_0$ with the only independent
renormalization constant $Z_{\nu}$:
\begin{equation}
\nu_0=\nu Z_{\nu}, \quad g_{0}=g\mu^{2\eps}Z_{g}, \quad
Z_{g}=Z_{\nu}^{-3} \quad (D_{0} = g_{0}\nu_0^{3} = g\mu^{2\eps} \nu^{3}).
\label{18}
\end{equation}
Here $\mu$ is the reference mass in the minimal subtraction (MS) scheme,
which we always use in what follows, $g$ and $\nu$ are renormalized analogs
of the bare parameters $g_{0}$ and $\nu_0$, and $Z=Z(g,\eps,d)$ are the
renormalization constants. No renormalization of the fields and the IR
scale $m_{0}=m$ is needed, i.e., $Z_{\Phi}=1$ for all $\Phi$ and
$Z_{m}=1$. The renormalized action has the form
\begin{equation}
S(\Phi)=v'D_{v}v'/2+v'\left[-\nabla_t+\nu Z_{\nu}\partial^{2}\right]v,
\label{Renact}
\end{equation}
where the amplitude $D_{0}$ in $D_{v}$ is expressed in renormalized
parameters using the last relation from (\ref{18}).

In the MS scheme the renormalization constants have the form ``1 + only poles
in $\eps$,'' in particular,
\begin{eqnarray}
Z_{\nu}=1+\sum _{k=1}^{\infty } a_k(g)\eps ^{-k}=1+\sum _{n=1}^{\infty }g^n
\sum _{k=1}^{n}a_{nk}\eps ^{-k},
\label{1.30}
\end{eqnarray}
where the coefficients $a_{nk}$ depend only on $d$.

Since the fields are not renormalized, their renormalized correlation
functions $G^{R}$ coincide with their unrenormalized analogs
$G=\langle\Phi\dots\Phi\rangle$; the only difference is in the choice of
variables and in the form of perturbation theory (in $g$ instead of $g_{0}$):
$G^{R} (g,\nu,\mu,m,\dots) = G (g_{0},\nu_0,m_{0},\dots)$. Here the dots
stand for other arguments like coordinates, times, momenta and so on. We
use $\Dm$ to denote the differential operator $\mu\partial_{\mu}$ for fixed
bare parameters $g_{0},\nu_0,m_{0}$ and operate on both sides of that
relation with it. This gives the basic differential RG equation:
\begin{equation}
{\cal D}_{RG}G^{R} (g,\nu,\mu,m,\dots) = 0, \quad
{\cal D}_{RG} = {\cal D}_{\mu} + \beta(g)\partial_{g}
-\gamma_{\nu}(g){\cal D}_{\nu},
\label{RGE}
\end{equation}
where ${\cal D}_{RG}$ is the operation $\Dm$ expressed in renormalized
variables, ${\cal D}_{x} = x\partial_{x}$ for any variable
$x$, and the RG functions (the anomalous dimension $\gamma_{\nu}$ and
the $\beta$ function) are defined as
\begin{eqnarray}
\gamma_\nu(g) = \Dm \ln Z_{\nu} = \beta(g,\varepsilon)
\partial_{g} \ln Z_{\nu},
\nonumber \\
\beta(g,\varepsilon) = \Dm g = g\left[-2\varepsilon+3\gamma_\nu(g)\right];
\label{RGF1}
\end{eqnarray}
the relation between $\beta$ and $\gamma_{\nu}$ results from the definitions
and the last relation in (\ref{18}). Combining the two relations in
(\ref{RGF1}) and substituting (\ref{1.30}) gives
\begin{equation}
\gamma_{\nu} (g)= -2 {\cal D}_{g}  a_1(g) + \ {\rm the\ terms\ containing\
only\ poles\ in\ }\ \eps.
\label{RGF22}
\end{equation}
with $a_1(g)$ from (\ref{1.30}). From the UV finiteness of the renormalized
functions $G^{R}$ it follows that the pole terms in (\ref{RGF22}) cancel
each other, which eventually gives
\begin{equation}
\gamma_{\nu}(g) = -2 {\cal D}_{g}  a_1(g) = -2 \sum _{n=1}^{\infty}\,
n\, a_{n1}\, g^{n}
\label{RGF23}
\end{equation}
with $a_1(g)$ and $a_{n1}$ defined in (\ref{1.30}). The cancellation
occurs due to the fact that the coefficients $a_{nk}$ with $k>1$
are algebraically related to $a_{n1}$, the coefficients in front of the
first-order poles, for example, $a_{22}=a_{21}^{2}$. Such relations can be
used to check the results of the practical calculations.

The one-loop result
\begin{eqnarray}
a_{11}=-(d-1)\bar S_{d}/8(d+2)
\label{a11}
\end{eqnarray}
with $\bar S_{d}$ from (\ref{surface}) is well known, while the two-loop
coefficient $a_{21}$ was calculated (for a few values of $d$, including
$d=3$ and in the limits $d\to2$ and $d\to\infty$) in \cite{APS}. In the
latter limit, which we are interested in here, one obtains:
\begin{eqnarray}
a_{21}= a_{11}^{2} \, \left\{ - \frac{1}{2} + O(1/d) \right\} .
\label{a21}
\end{eqnarray}

From the first order result (\ref{a11}) it follows that the $\beta$ function
has a nontrivial fixed point [$\beta(g_*)=0$] in the physical region $g>0$
with the coordinate
\begin{equation}
g_* = 8(d+2)\eps\, /\, 3(d-1)\bar S_{d} +O(\eps^{2}).
\label{FP1}
\end{equation}
The correction exponent $\omega =\beta'(g_*)=2\eps +O(\eps^{2})>0$ at this
point is positive, so that it is IR attractive and governs the IR behaviour
of the correlation functions. The value of $\gamma_{\nu}(g)$ at the fixed
point is found exactly using the relations (\ref{RGF1}):
\begin{equation}
\gamma_{\nu}^{*} \equiv \gamma_{\nu}(g_*)= 2\eps/3.
\label{FPE}
\end{equation}

The relations (\ref{a11}) and (\ref{a21}) illustrate the fact that the
coefficients that have a well-defined finite limit for $d\to\infty$ are
$b_{n1} =  a_{n1} \,\bar S_{d}^{-n}$ rather than $a_{n1}$ themselves
(this is obvious from the practical calculation, see the next
section~\ref{sec:Grin}). It is convenient to introduce the new coupling
constant by the relation
\begin{eqnarray}
u = g \bar S_{d} .
\label{u}
\end{eqnarray}
Then the correlation functions $G$ and $G^{R}$, renormalization constants
(\ref{1.30}) and the RG functions $\gamma_\nu$ and  $\beta= \Dm u$ will
have well-defined finite limits $d\to\infty$ when expressed as perturbation
series in the parameter $u$; the latter is kept finite in that limit.

In the following, the large-$d$ limit will always be understood in this
sense. From the physics viewpoints it corresponds to the natural requirement
that the quantity which is kept finite for $d\to\infty$ is the energy input
per one component of the velocity field (that is, per one spatial
dimension), which is clear from the relations (\ref{surface}), (\ref{2.74})
and (\ref{Prad}). Such a choice is also consistent with further RG analysis
which shows that the fixed-point value $u_{*}$ of the parameter $u$
(and not that of $g$) is finite for $d\to\infty$, as illustrated by
equation (\ref{FP1}).

\section{Calculation of the renormalization constant and RG functions
to the third order} \label{sec:Grin}

Let us turn to the calculation of the renormalization constant $Z_{\nu}$
in (\ref{1.30}) with the accuracy $O(g^{3})$ (three-loop approximation)
in the limit $d\to\infty$. In general, the renormalization constant
$Z_{\Gamma}$ corresponding to a certain 1-irreducible Green function
$\Gamma=\langle \Phi\dots\Phi\rangle_{\rm 1-ir}$ can be found from
the relation \cite{Vladim} (see also the monograph \cite{Book3})
\begin{eqnarray}
Z_{\Gamma} = 1 - {\cal KR'} \tilde \Gamma,
\label{Vlad}
\end{eqnarray}
where $\tilde \Gamma$ is the function $\Gamma$ normalized with respect to
unity in the zeroth order of the perturbation theory, ${\cal R'}$ is the
incomplete ${\cal R}$-operation which involves all the subtractions for
the divergent subgraphs, without the last subtraction for the diagram as
a whole, and ${\cal K}$ is the subtraction operation for the given
renormalization scheme. In the MS scheme, which we use in our calculation,
${\cal K}$ subtracts only poles in $\eps$:
\[ {\cal K} \sum_{n=-\infty}^{\infty} a_{n}\eps^{n} =
\sum_{n=-\infty}^{-1} a_{n}\eps^{n} \]
for any Laurent series. The function $\tilde \Gamma$ in (\ref{Vlad}) should
be calculated in terms of the renormalized parameters from the ``basic''
action functional \cite{Book3,Collins} which is obtained from the
renormalized action (\ref{Renact}) by the replacements $Z_{i}\to 1$ for all
the renormalization constants.

The only independent renormalization constant $Z_{\nu}$ in our model
(\ref{Renact}) is determined by the 1-irreducible function (in the
frequency-momentum representation)
\begin{equation}
\Gamma_{ij}(\omega,{\bf p})= \langle v'_{i}v_{j} \rangle_{\rm 1-ir} =
\Gamma(\omega,{\bf p}) P_{ij}({\bf p}),
\label{Ra}
\end{equation}
where the projector $P_{ij}({\bf p}) = \delta _{ij} - p_i p_j /p^2$
arises due to the transversality of the fields $v'_{i}$, $v_{j}$.
Thus the scalar coefficient $\Gamma(\omega,{\bf p})$ in (\ref{Ra}) is given
by the relation
\begin{equation}
\Gamma(\omega,{\bf p}) = \Gamma_{ij}(\omega,{\bf p})P_{ij}({\bf p}) / (d-1)
= \Gamma_{ii}(\omega,{\bf p}) / (d-1),
\label{RaP}
\end{equation}
where $(d-1) = P_{ii}({\bf p})$ comes from the trace of the projector.
In the perturbation theory,
\[ \Gamma(\omega,{\bf p}) =  {\rm i}\omega -\nu p^{2}+O(g). \]
Since the counterterm to the function (\ref{Ra})
is proportional to $\nu p^2$, one can put $\omega=0$ in $\Gamma$
and neglect the higher-order terms in $p^{2}$. Therefore the normalized
scalar function $\tilde\Gamma$ in (\ref{Vlad}) can be taken as
\begin{eqnarray}
\tilde\Gamma= \lim_{{\bf p}\to 0}
\frac{\bigl\langle v'_{i} v_{i} \bigr\rangle_{\rm 1-ir} (\omega=0;{\bf p})}
{\nu p^2 (1-d)} = 1+ O(g);
\label{Ratio}
\end{eqnarray}
it depends only on the completely dimensionless variables $g$ and $m/\mu$.

The elementary integrations over times (or equivalently frequencies) in the
diagrams are always easily performed; the serious problem is only the
integration over the momenta. (The factors $\nu$ arising from these
integrations and from the propagator $\langle vv \rangle_{0}$
group together to cancel $\nu$ in the denominator of (\ref{Ratio}).)

The first important observation is that, for any diagram of the function
(\ref{Ra}), the number of loops (that is, the number of independent
integration momenta) is equal to the number of the lines (propagators)
of the type $\langle vv \rangle_{0}$ (this is easy to understand from
a few examples, given in this section, so that we shall not give the
formal proof). It is therefore possible to assign
the ``pure'' independent momenta (which we shall denote by ${\bf k}$,
${\bf q}$ and ${\bf l}$ for the three-loop diagrams) to these propagators.
Then the momenta associated with the remaining lines of the type
$\langle vv' \rangle_{0}$ will be certain definite linear combinations of
the external momentum argument ${\bf p}$ and the integration momenta
${\bf k}$, ${\bf q}$ etc.

Each line $\langle vv \rangle_{0}$ brings about one integration over an
independent pure momentum:
\begin{equation}
\int \frac{d{\bf k}}{(2\pi)^{d}} \langle vv \rangle_{0} ({\bf k}) \dots =
\frac{g\mu^{2\eps}\nu^2}{(2\pi)^{d}} \int d{\bf k} k^{2-d-2\eps} \dots .
\label{Int1}
\end{equation}
Here we used the last expression from (\ref{linesV}) and omit the
dimensionless time-dependent factor and the projector, denoted by the
ellipsis. Now we can separate the integration over ${\bf k}$ into the
integrations over the modulus $k=|{\bf k}|$ and the direction
${\bf n}={\bf k}/k$ (``angles''), which gives
$d{\bf k} = k^{d-1}dk d{\bf n}$, and replace the integration over
the angles by the averaging $\langle\dots\rangle_{S}$ over the unit
$d$-dimensional sphere, normalized such that $\langle 1\rangle_{S}=1$.
This gives
\begin{equation}
\frac{g\mu^{2\eps} \nu^2}{(2\pi)^{d}} \int_{m}^{\infty}
{dk}\, {k^{1-2\eps}} \int d{\bf n} \dots = g\mu^{2\eps}
\nu^2 \bar S_{d} \int_{m}^{\infty} {dk}\, {k^{1-2\eps}} \
\langle\dots\rangle_{S}.
\label{Int2}
\end{equation}
The factor $\bar S_{d}$ from (\ref{surface}) groups together with the
coupling constant $g$; this quantity is kept fixed in the limit $d\to\infty$.
The lower limit $m$ in the integration over $k$ comes from (\ref{1.99}) and
provides the IR regularization.

In (\ref{Int2}) the powers $k^{-d}$ from the correlator (\ref{1.9}) and
$k^{d}$ from the Jacobian cancel each other, so that the dependence on $d$
in the exponent disappears. The remaining dependence on $d$ comes from the
contractions of the projectors (\ref{linesV}) and vertices (\ref{vertexV}),
with further integrations over the angles.

The vertex $V_{ijs}^{S}$ in (\ref{vertexV}) is explicitly symmetric with
respect to the indices $js$, that is, symmetric with respect to the two
attached fields $v_{j}$ and $v_{s}$. Let us split it into the two asymmetric
parts
\begin{equation}
V_{ijs}^{S}=V_{ijs}^{A}+V_{isj}^{A},
\label{vertexRR}
\end{equation}
where the new asymmetric vertex
$V_{ijs}^{A}= {\rm i} k_{j}\delta_{is}$ in the diagrams will be denoted as
\begin{equation}
V_{ijs}^{A} = {\rm i} k_{j}\delta_{is} = - {\rm i} p_{j}\delta_{is} = \
\raisebox{-0.40cm}{\includegraphics[height=1.3cm,width=1.8cm]{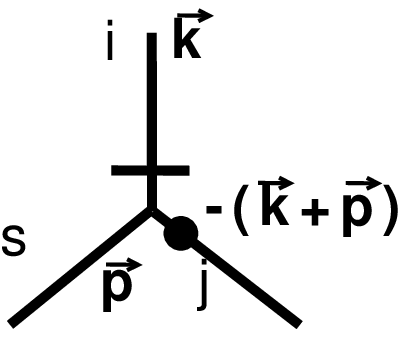}},
\label{vertexA}
\end{equation}
so that the splitting (\ref{vertexRR}) is diagrammatically represented as
\begin{equation}
V_{ijs}^{S} =
\raisebox{-0.40cm}{\includegraphics[height=1.0cm,width=4.0cm]{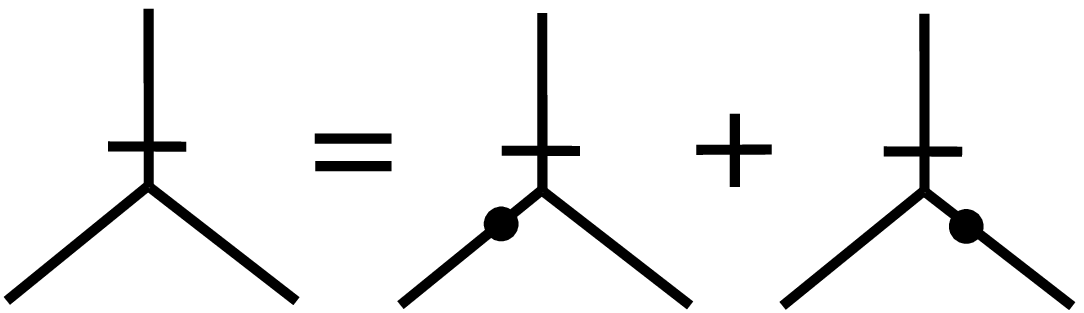}}.
\label{spilt}
\end{equation}
Here, the end marked with the slash corresponds to the field $v'$, the end
with no marks corresponds to the field $v$ which stands under the
derivative in the expression $-v'(v\partial)v= v'_{i}V^{A}_{ijs}v_{j}v_{s}$;
the vector indices of these two fields are contracted to each other.
The end marked with the thick dot corresponds to the second field $v$
(without a derivative); its vector index is contracted with the index of
the momentum. The momentum
${\bf k}$ is the argument of the field $v'$. In the asymmetric vertex, it
can be equally replaced with $-{\bf p}$, the argument (up to the minus sign)
of the unmarked field $v$, due to the transversality of the marked field $v$.

The use of the asymmetric vertex leads, of course, to increase in the
number of diagrams: each diagram with $N$ symmetric vertices (\ref{vertexV})
gives rise to the sum of $2^{N}$ contributions, represented by diagrams
with vertices (\ref{vertexA}), as illustrated in figure~1 for the one-loop
case. This inconvenience is compensated by the fact that the large-$d$
behaviour of such individual contributions is easier to analyze.
We shall see below that most of them vanish in the limit $d\to\infty$,
and it is possible not to draw the corresponding diagrams from the very
beginning.

\begin{figure}
\begin{center}
\includegraphics[width=12cm]{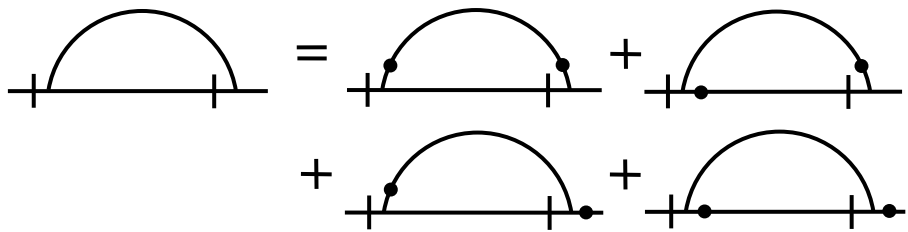}
\caption{Passage to the asymmetric vertex in the one-loop approximation.}
\end{center}
\end{figure}

From (\ref{RaP}) it follows that a diagram can survive in the limit
$d\to\infty$ only if its contribution to the quantity
$\Gamma_{ii}(\omega,{\bf p})$ in the numerator behaves as $O(d)$ for large
$d$. The only possible source of such a contribution is a closed
(self-contracted) chain of intermittent $\delta$-symbols,
\begin{equation}
\delta_{ii_{1}} \delta_{i_{1}i_{2}} \dots \delta_{i_{n-1}i_{n}}
\delta_{i_{n}j} \delta_{ji},
\label{Chain}
\end{equation}
coming from the vertices (\ref{vertexA}) and from the projectors in
propagators (\ref{linesV}), with the final contraction with $\delta_{ji}$
from the projector in (\ref{RaP}). It is easily seen that, in terms of the
asymmetric vertex, each diagram can involve no more than only one
chain of the type (\ref{Chain}). Assume for definiteness that
the factor $\delta_{ii_{1}}$ in (\ref{Chain}) comes from the leftmost
vertex of the diagram, then $\delta_{i_{n}j}$ comes from the rightmost
vertex. Then from the form of the vertex (\ref{vertexA}) it follows that
the vector index $i$ corresponds to the field $v$ with a derivative,
and $i_{1}$ corresponds to the field $v'$. The latter is necessarily
contracted with the field $v$ in the neighboring vertex, because the
propagator $\langle v'v'\rangle_{0}$ in (\ref{linesV}) vanishes
identically. Thus the next factor $\delta_{i_{1}i_{2}}$ comes from the
propagator $\langle v'v\rangle_{0}$. It is contracted with the next symbol
$\delta_{i_{2}i_{3}}$ from the vertex, in which $i_{2}$ must correspond to
the field $v$ with a derivative (any contraction with the momentum
in (\ref{vertexA}) would break the chain of $\delta$ symbols); then
$i_{3}$ necessarily corresponds to the field $v'$, and so one.

We may conclude that the product of the type (\ref{Chain}) is necessarily
associated with the chain of propagators $\langle v'v\rangle_{0}$ which
begins in the leftmost vertex of the diagram and ends up in the rightmost
vertex; we shall call it ``the spine'' in the following.\footnote{The factor
of the form (\ref{Chain}) can also be associated with a {\it closed} circuit
of the propagators $\langle v'v\rangle_{0}$, not connected to the external
vertices of the diagram.
But such diagrams vanish identically owing to the fact the propagator
$\langle v'v\rangle_{0}$ is retarded: it contains the step function in the
time arguments; see (\ref{linesV}).} It can be formally proved, but
is easily seen from the examples, that each diagram of the 1-irreducible
function (\ref{Ra}) involves one and only one spine. However, the spine
gives rise to the factor of the type (\ref{Chain}) only if all the fields
$v$ enter it with derivatives, so that their vector indices are contracted
with the indices of the fields $v'$; this means that all the dots in the
vertices (\ref{vertexA}) must be placed out of the spine. This fixes the
placements of the dots in the vertices that belong to the spine in the
unique way and drastically reduces the number of relevant diagrams. In
figure~1, only the first diagram contains the factor (\ref{Chain}) and
gives a nonvanishing contribution to the quantity (\ref{RaP});
the other diagrams vanish for $d\to\infty$ and can be dropped from the
very beginning. Figure~2 gives examples of nonvanishing three-loop diagrams
with the proper placement of the dots on the spine
(here and in the figures below, the spine is always shown as a horizontal
chain of propagators).

\begin{figure}
\begin{center}
\includegraphics[width=12cm]{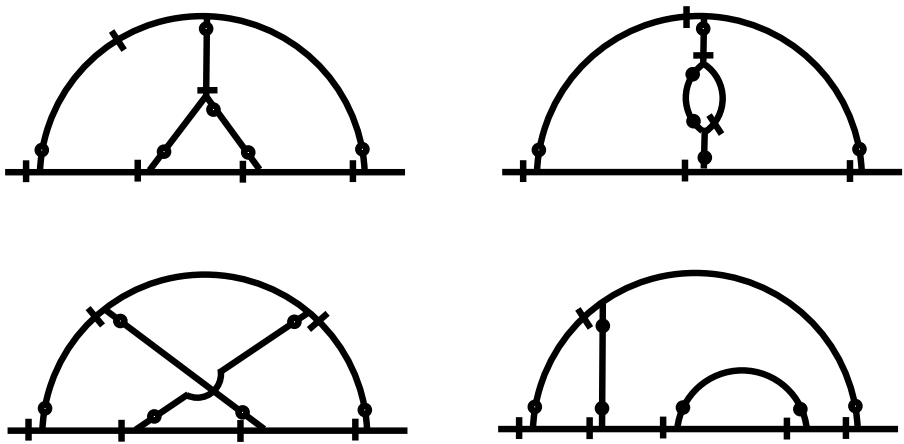}
\caption{Examples of three-loop diagrams which do not
vanish for $d\to\infty$.}
\end{center}
\end{figure}

The next important observation is that any nontrivial angular integration
brings about smallness in $1/d$; the corresponding contributions in the
diagrams can be dropped for $d\to\infty$.
More precisely, the average over the unit
sphere of the scalar product of two independent momenta ${\bf k}$ and
${\bf q}$ with the angle $\vartheta$ between them is given by the relations
\begin{eqnarray}
\langle({\bf kq})^{2n}\rangle_{S} = (kq)^{2n} \langle \cos^{2n}
\vartheta \rangle_{S} =
\nonumber \\
= (kq)^{2n}\, \frac{(2n-1)!!} {d(d+2)\dots(d+2n-2)} =O(d^{-n})
\label{L2}
\end{eqnarray}
and $\langle({\bf kq})^{2n+1}\rangle_{S}=0$; here $n=1,2$ and so on.
In the three-loop diagrams, averages of the form
\begin{equation}
\langle ({\bf kq})^{n_{1}}({\bf kl})^{n_{2}}({\bf lq})^{n_{3}} \rangle_{S}
\label{L3}
\end{equation}
arise in the integrands. The corresponding expressions, analogous to
(\ref{L2}), are rather complicated; they can be found e.g. in sec.~VIC
of Ref.~\cite{RG3}. It is only important for us here, that they always
vanish for $d\to\infty$ (with the obvious trivial exception $n_{i}=0$
for all $i$.)

In all the relevant diagrams, the momentum argument of the two
external vertices can be chosen equal to ${\bf p}$, the external
momentum argument of the function (\ref{Ra}). [This would be
impossible for the diagrams in which the dot in the rightmost
external vertex is placed on the external line, like in the last
two diagrams in figure~1. The momentum arguments for such vertices
necessarily involve internal integration momenta ${\bf k}$, ${\bf q}$
and so one. But we have already established that for
$d\to\infty$ their contributions vanish.] Thus the overall factor
of the form $p_{i}p_{j}$ is isolated from the quantity
(\ref{RaP}), and, since we are interested in the limit (\ref{Ra}),
we can set ${\bf p}=0$ in the rest of the integrand.

If one of the external momenta, say $p_{i}$, is contracted with some
integration momentum in the integrand, say $k_{i}$ or $q_{i}+k_{i}$,
a construction with scalar products of the form (\ref{L2}) or (\ref{L3})
arises, and in the limit $d\to\infty$ the diagram vanishes after the
angular integrations. Nonvanishing contribution arises only if the
external momenta are contracted with each other through a chain
of $\delta$ symbols:
\begin{equation}
p_{i} \delta_{ii_{1}} \delta_{i_{1}i_{2}} \dots \delta_{i_{n-1}i_{n}}
\delta_{i_{n}j} p_{j}.
\label{Chain2}
\end{equation}
In the diagram, it corresponds to the chain of propagators that connects the
two external vertices, with only two dots placed on the chain at the external
vertices; they correspond to the external momentum. Each dot placed on the
chain at an internal vertex corresponds to contraction with some integration
momentum ${\bf k}$, ${\bf q}$ etc, and therefore breaks the chain
(\ref{Chain2}). All the diagrams shown in figure~2 have such ``second spine,''
having the form of a big arc connecting the external vertices and surrounding
the diagram from above; the proper placement of the dots is also shown. From
these examples it is clear that the second spine always has the form
\begin{equation}
\langle vv'\rangle_{0}\langle vv'\rangle_{0} \dots \langle vv'\rangle_{0}\
\langle vv\rangle_{0}\ \langle v'v\rangle_{0} \dots \langle v'v\rangle_{0},
\label{Arc}
\end{equation}
that is, it consists of two chains of mixed propagators
$\langle vv'\rangle_{0}$ and $\langle v'v\rangle_{0}$ (not necessarily of
equal ``lengths''), proceeding from the opposite directions and connecting
with the only propagator $\langle vv\rangle_{0}$. In the simplest case of
the only nonvanishing one-loop diagram in figure~1, the ``second spine''
exists and consists of the single propagator $\langle vv\rangle_{0}$.

If such ``second spine'' exists, it is unique because the chain of $\delta$
symbols in (\ref{Chain2}) ``cannot branch.'' But the existence is not
guaranteed (in contrast to the ``first spine'' discussed above), as
illustrated by the diagrams in figure~3. The first spines,
shown as horizontal lines, consist of 5 and 3 propagators, respectively
(the proper placement of the dots, which gives rise to the
chains of the form (\ref{Chain}), is always unique and is not shown).
But the second spines do not exist: for the first diagram, there is
no chain of propagators, connecting the external vertices. For the
second diagram, such a chain exists (the big arc) but it involves {\it two}
propagators $\langle vv\rangle_{0}$ and thus it does not have the form
(\ref{Arc}); any placement of the dot in the right
upper vertex breaks the chain of $\delta$ symbols in (\ref{Chain2}).

Such diagrams give no contribution for $d\to\infty$, so
they can be dropped from the very beginning. In the nonvanishing diagrams,
the factor $p^{2}$ coming from the chain (\ref{Chain2}), explicitly cancels
the analogous factor in the denominator of (\ref{Ratio}); the remaining
quantity depends only on the dimensionless parameters $g$, $\eps$ and
$\mu/m$ (the latter will also disappear soon).

\begin{figure}
\begin{center}
\includegraphics[width=12cm]{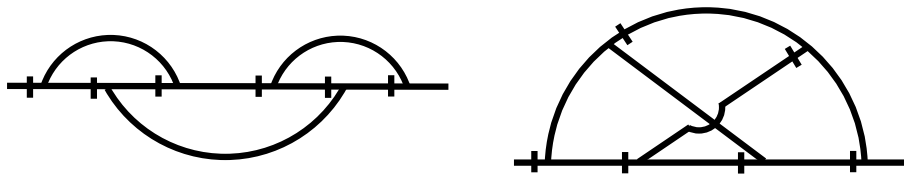}
\caption{Examples of three-loop diagrams which have no ``second spine''
and vanish for $d\to\infty$.}
\end{center}
\end{figure}

Discarding the diagrams which vanish for $d\to\infty$ (those with ``wrong''
placement of the dots on the spine or those having no second spine) leads
to impressive decrease in the number of the surviving diagrams. In the
one-loop approximation only one of four diagrams survives (as is clear from
figure~1), in the two-loop approximation 6 diagrams of 120 survive, and
in the three-loop approximation only 80 diagrams survive of 8160.

The final simplification in the remaining nonvanishing diagrams is the
possibility to drop all the scalar products of the integration momenta
${\bf k}$, ${\bf q}$ and so on, due to the smallness in the averages in
expressions (\ref{L2}) and (\ref{L3}).  If a scalar product appears in
the denominator of the integrand, say $k^{2}+q^{2}+({\bf kq})$, these
expressions are not applied directly, but one can expand all the
denominators in all the scalar products and then discard all the terms
except for the first one (with no scalar products) due to relation
(\ref{L2}) and its analog for more complicated structures like (\ref{L3}).
This is the same as to drop all the scalar products in the integrals from
the very beginning. The diagrams in which the integrands are proportional
to some scalar products
vanish after this procedure, but we have no general rule how to identify
them from the very beginning. The averaging over the angles now becomes
trivial: $\langle 1\rangle_{S}=1$.

Thus the expressions for the remaining nonvanishing diagrams in the $n$-loop
approximation are represented by $n$-fold integrals over the $n$ scalar
variables, the moduli of the independent integration momenta $k$, $q$
and so one. It is convenient to pass to new integration variables $k^{2}$,
$q^{2}$ and so one; then $m^{2}$ will appear as the new IR cutoff in the
integrals. We shall retain the same notation $k$, $q$, $m$ for these new
parameters. Then, for $n=3$, such integrals take on the forms
\begin{equation}
J(m) = \int_{m}^{\infty} \frac{dk}{k^{1+\eps}}\int_{m}^{\infty}
\frac{dq}{q^{1+\eps}} \int_{m}^{\infty} \frac{dl}{l^{1+\eps}} \,
f(k,q,l),
\label{IntS}
\end{equation}
where $f$ is some dimensionless function. From dimensionality
considerations it is clear that the integral (\ref{IntS}) is the definite
power of the IR scale, $J(m)= m^{-3\eps} J(1)$. Then applying the
differential operation $\D_{m}=m\partial/\partial m$ to the integral
(\ref{IntS}) multiplies it by the factor $(-3\eps)$, and the following
identity holds:
\begin{equation}
J(1) = -\frac{1}{3\eps} \D_{m} J(m)|_{m=1}.
\label{Iden}
\end{equation}
The parameter $m$ enters the expression (\ref{IntS}) only in the lower
integration limits, so that explicit differentiation in the right-hand side
of (\ref{Iden}) reduces the number of integrations by one. Since one pole
in $\eps$ in (\ref{Iden}) is already isolated explicitly, in the
resulting double integrals one should retain the terms of order
$\eps^{-2}$, $\eps^{-1}$ and $\eps^{0}$.

The complete three-loop calculation involves only six basic integrals
$J_{i}$ of the form (\ref{IntS}) with the following functions $f_{i}$:
\begin{eqnarray}
f_{1} &=& \frac{kql}{(k+q+l)^{3}}, \quad
f_{2} = \frac{kq}{(k+q+l)^{2}}, \quad
f_{3} = \frac{kql}{(k+q)(k+l)(q+l)}, \nonumber \\
f_{4} &=& \frac{kq}{(k+q)(k+q+l)}, \quad
f_{5} = \frac{kq}{(k+q)(k+l)}, \quad f_{6}= \frac{kq}{(k+q)^{2}}.
\label{fis}
\end{eqnarray}
The practical calculation gives
\begin{eqnarray}
\tilde J_{1} &=& \frac{1}{2}, \quad
\tilde J_{2} = \frac{1}{2\eps}, \quad
\tilde J_{3} = \frac{\pi^2}{4}, \quad
\tilde J_{4} = \frac{1}{2\eps}+1, \nonumber \\
\tilde J_{5} &=& \frac{1}{2\eps^{2}} + \frac{\pi^2}{6}, \quad
\tilde J_{6} = \frac{3}{2\eps} -3 \ln2,
\label{Jis}
\end{eqnarray}
where we denoted $J_{i}(1) = (1/3\eps) \tilde J_{i}$.

The ${\cal R}$-operation in (\ref{Vlad}) implies subtraction of the UV
divergences related to the subgraphs; this is a standard routine,
see e.g. \cite{Book3,Collins}. It requires calculation of the one-loop
and two-loop analogs of the integrals like (\ref{IntS}) with the proper
accuracy; this is of course an easier task than the calculation of the
three-loop integrals. Combining all the contributions finally gives the
following third-order result for the renormalization constant (\ref{1.30}):
\begin{equation}
Z_{\nu} = 1 - {\displaystyle \frac {u}{8\,\varepsilon }}  -
\frac{u^{2}\left(2+\eps\right)}{128\eps^{2}}  -
\frac{u^{3}(10+9\eps+7\eps^{2})}{3072\eps^{3}} +O(u^{4}) ,
\label{Znu3}
\end{equation}
with the coupling constant $u$ introduced in (\ref{u}). It is worth noting
that the transcendental numbers ($\ln 2$ and $\pi^2$) entering the
expressions (\ref{Jis}) have disappeared in (\ref{Znu3}) as a result of
the subtraction of subdivergences.

Then for the anomalous dimension (\ref{RGF1})
using the relation (\ref{RGF22}) one finds:
\begin{equation}
\gamma_{\nu}(u) = {\displaystyle \frac {1}{4}} \,u + {\displaystyle
\frac {1}{32}} \,u^{2} + {\displaystyle \frac {7}{512}} \,u^{3} +O(u^{4})
\label{gamma3}
\end{equation}
and the three-loop $\beta$ function is found by the second relation in
(\ref{RGF1}). Solving the equation $\beta(u_{*})=0$ perturbatively in $\eps$,
for the coordinate of the fixed point one obtains:
\begin{equation}
u_{*}= {\displaystyle \frac {8}{3}} \,\varepsilon  -
{\displaystyle \frac {8}{9}} \,\varepsilon ^{2} - {\displaystyle
\frac {4}{9}} \,\varepsilon ^{3} + O(\eps^{4}).
\label{ufix3}
\end{equation}
The UV correction exponent
\begin{equation}
\omega = \beta'(u_{*}) =  2\,\varepsilon  + {\displaystyle \frac {2}{3}} \,
\eps^{2} + {\displaystyle \frac {10}{9}} \,\eps ^{3} + O(\eps^{4})
\label{omega3}
\end{equation}
completes the list of the third-order results.

The expressions obtained appear rather simple: in particular, all
the coefficients in the series in $g$ or $\eps$ for $Z_{\nu}$, $\beta$,
$u_{*}$ and $\omega$ are rational numbers, although for dynamical models
such coefficients usually involve transcendental quantities (logarithms,
hypergeometric functions etc) already in the two-loop approximations.
One can hope that this fact remains valid for all the higher-order terms.
It is then tempting to try to guess their general form and, therefore,
to construct some hypothetical exact (that is, beyond the expansion in $g$
or $\eps$) expressions for those quantities. Such exact answers for all the
quantities (\ref{Znu3})--(\ref{omega3}) exist \cite{Xess} for the well-known
Heisenberg model of the developed turbulence \cite{Heis,Monin}. In
particular, for $\omega$ and $u_{*}$ (in the MS scheme) one has \cite{Xess}:
\begin{equation}
\omega= \eps (1-\eps/3) /(1-5\eps/12), \quad
u_{*} = (\eps/3) (1-\eps/3)^{1/4},
\label{omegaH}
\end{equation}
where all the parameters have the same meaning as in our model (it is
implied, of course, that the usual Heisenberg model is generalized to
the energy pumping of the form (\ref{1.10})).  Let us assume that in our
model the fixed-point coordinate has a similar form:
\[ u_{*} = (c_{1}\eps) (1+\eps c_{2})^{c_{3}}. \]
Then comparison with the cut (\ref{ufix3}) of the $\eps$ expansion for
$u_{*}$ allows one to find all the coefficients $c_{i}$, which appear
very simple and, surprisingly enough, the exponent $c_{3}=1/4$ is exactly
the same as in the Heisenberg model:\begin{equation}
u_{*} = (8\eps/3) (1-4\eps/3)^{1/4}.
\label{omegaE}
\end{equation}
Now the exact expression for $\omega$ can be find from (\ref{omegaE}).
Differentiating the equation (\ref{RGF1}) for the function
$\beta =\Dm u$ with respect to $u$, setting $u=u_{*}$ and taking into
account $u_{*}\ne0$ gives:
\[ \omega = \beta'(u_{*}) = 3 \gamma_{\nu}'(u_{*}), \]
and differentiating the equation (\ref{FPE}) with respect to $\eps$ gives:
\[ \gamma_{\nu}'(u_{*}) u_{*}'(\eps)=2/3 \]
(the first differentiation with respect to $u$ and the second -- with
respect to $\eps$). Combining these equalities gives the exact relation
\[ \omega(\eps) = 2u_{*}/u_{*}'. \]
Substituting the explicit expression (\ref{omegaE}) for the function
$u_{*}=u_{*}(\eps)$ gives the desired hypothetical exact answer for the
UV correction exponent:
\begin{equation}
\omega= 2\eps(1-4\eps/3) /(1-5\eps/3),
\label{omegaEE}
\end{equation}
which looks very similar to the corresponding expression (\ref{omegaH}) for
Heisenberg's model and coincides (as is easily checked) with the
expansion (\ref{omega3}). From (\ref{omegaE}) and (\ref{omegaEE}) it is
then possible to derive the exact expressions for the $\beta$ function,
anomalous dimension $\gamma_{\nu}$ and the renormalization constant
$Z_{\nu}$; cf. \cite{Xess} for the Heisenberg model and \cite{Pavel}
for the Kraichnan model. These expressions are rather cumbersome and will
not be reported here.

Expressions (\ref{omegaE}), (\ref{omegaEE}), provided they are correct,
show that the IR fixed point $u_{*}>0$ disappears at $\eps=3/4$, that is,
before the physical value $\eps=2$ is achieved. This must lead to some
qualitative change in the IR behaviour of the correlation functions
which cannot be properly described within any finite-order approximation
of the $\eps$ expansion. Of course, one should not take too seriously these
hypothetical expressions, which, of course, are not the only possible ones.

\section{Pair correlation function: Third-order approximation}
\label{sec:pair}

In this section we shall discuss the equal-time pair correlation function
of the velocity field in the momentum representation,
\begin{equation}
D_{ij}({\bf p})=P_{ij}({\bf p})\, D(p).
\label{G}
\end{equation}
We are ultimately interested in its inertial-range behaviour ($m\ll p$),
so in the following we set $m=0$; the IR regularization in the diagrams
is provided by the momentum $p$. Solving the RG equation (\ref{RGE})
for the function $D(p)$ gives (see e.g. \cite{Book3,UFN,turbo} for the
detailed derivation)
\begin{equation}
D(p) =  g \nu^2 p^{-d+2} R(s,g) \simeq D_{0}^{2/3} g_{*}^{1/3}
p^{-d+2\Delta_{v}} R(1,g_{*}), \quad s = p/\mu.
\label{dimG}
\end{equation}
The first equality, along with dimensionality considerations, introduces the
function $R$ which depends on the two dimensionless variables $g$ and
$s = p/\mu$ (dependence on $\eps$ and $d$ is always implied).
The second relation
holds in the IR asymptotic region $s = p/\mu \ll 1$ and involves the
coordinate of the IR attractive fixed point $g_{*}$ and the critical
dimension of the velocity field, which is known exactly:
$\Delta_{v}=1-2\eps/3$. It depends only on the amplitude $D_{0}$ from
(\ref{1.10}) and is independent of the viscosity coefficient $\nu_0$
in agreement with the second Kolmogorov hypothesis; see the discussion
of this issue in \cite{Book3,UFN,turbo,JETP}.

The function $R$ can be directly calculated in the renormalized perturbation
theory as a series in $g$, its coefficients being finite at $\eps\to0$.
[One factor $g$ in (\ref{dimG}) is explicitly isolated from $R$ such that
its expansion in $g$ begins with $O(g^{0})$.] Substituting $g_{*}$ as the
series in $\eps$ (with the first-order term given by (\ref{FP1})) gives
the $\eps$ expansion of the amplitude $R(1,g_{*})$.
It was calculated earlier in the one-loop approximation for certain values
of $d$; see \cite{APS}. Here we shall calculate it in the limit $d\to\infty$
in the two-loop approximation (three terms of the $\eps$ expansion). This
accuracy is consistent with the three-loop calculation of the
renormalization constants and the coordinate $g_{*}\sim u_{*}$,
presented in section~\ref{sec:Grin}.

The renormalized perturbation theory for the function (\ref{G}) has the form
\begin{equation}
D(p) = \frac{g\nu^{2}}{2 p^{\,d-2}}\,
\left(\frac{\mu}{p}\right)^{2\eps} \left\{1+ c_{1}g
\left(\frac{\mu}{p}\right)^{2\eps} + c_{2}g^{2}
\left(\frac{\mu}{p}\right)^{4\eps} + \dots \right\}
\label{G2}
\end{equation}
with some dimensionless coefficients $c_{i}=c_{i}(\eps,d)$. The first-order
term is the bare propagator $\langle v_{i}(t) v_{j}(t') \rangle _0$ from
(\ref{linesV}), in which we set $t=t'$ and replaced $\nu_0\to\nu$ and so on.
The contribution $D^{(n)}$ of a certain $n$-loop diagram to the series
(\ref{G2}) has the form
\begin{equation}
D^{(n)}(p)= \frac{g^{n+1} \nu^{2}  A^{(n)}} { p^{\,d-2}}
\left(\frac{\mu}{p}\right)^{2\eps(n+1)},
\label{G3}
\end{equation}
the problem is to calculate the dimensionless coefficient
$A^{(n)}=A^{(n)} (\eps,d)$. After the simple integrations over the time
variables, the quantity (\ref{G3}) is represented as the integral over $n$
independent momenta.

In contrast to the Green function (\ref{Ra}), now it is impossible,
in general, to assign these momenta to all the propagators
$\langle vv\rangle_0$.
The problem arises if the diagram includes a 1-irreducible subgraph of the
type $\langle v'v'\rangle_{\rm 1-ir}$, which itself is not a subgraph of
another 1-irreducible subgraph (then it depends as a whole on the external
momentum ${\bf p}$). Then, as illustrated by figure~4, some of the
propagators $\langle vv\rangle _0$ necessarily correspond to certain linear
combinations of the independent momenta and the external momentum $\p$,
and not to a ``pure'' independent momentum. Thus the cancellation of the
powers like $k^{-d}$ from the correlator (\ref{1.9})
and $k^{d}$ from the Jacobian (discussed for the Green function below the
equation (\ref{Int2})) does not occur, the dependence on $d$ in the exponent
does not disappear, and the calculational techniques for $d\to\infty$,
developed in the previous section, are not directly applicable.

\begin{figure}
\begin{center}
\includegraphics[width=12cm]{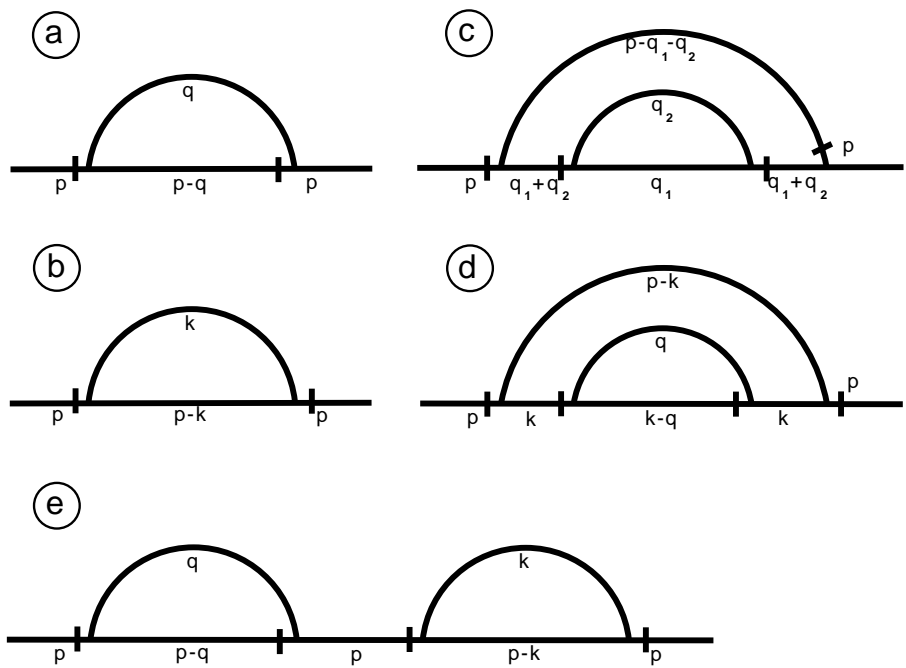}
\caption{Examples of the one-loop and two-loop diagrams for the
equal-time pair correlation function.}
\end{center}
\end{figure}

The general situation is illustrated by specific examples given in figure~4.
The external momentum is denoted by $\p$ and the independent integration
momenta are ${\bf k}$, ${\bf q}$ and ${\bf q}_{1,2}$ (the difference between
``${\bf k}$-momenta'' and ``${\bf q}$-momenta'' will become clear a bit
later). The problem of the non-cancellation does not occur for the diagrams
{\it a} and {\it c}: the diagram {\it a} has no subgraphs of the type
$\langle v'v'\rangle_{\rm 1-ir}$, while the diagram {\it c} has one such
subgraph, but it belongs to another subgraph. Thus it is possible to assign
the pure integration momenta (denoted by ${\bf q}$) or the external momentum
${\bf p}$ to all the $\langle vv\rangle_{0}$ lines, as shown in the figure.
The diagrams {\it b}, {\it d} and {\it e} involve one ``harmful'' subgraph
of the form $\langle v'v'\rangle_{\rm 1-ir}$, and one of the
$\langle vv\rangle_{0}$ lines necessarily corresponds to certain linear
combination of ``pure'' momenta; in all the examples, it is
${\bf p}-{\bf k}$. These examples also illustrate the general fact that
any diagram can involve no more than one  ``harmful''
$\langle v'v'\rangle_{\rm 1-ir}$ subgraph. What is more, as illustrated
by the diagram {\it e}, the problem of the non-cancellation occurs only
within this subgraph; the other subgraphs involve only momenta of the
type ${\bf q}$.

In principle, the problem can be attacked by the steepest descent method,
but in practice it is hardly possible to find the stationarity manifold
in the integration region.
In order to circumvent this difficulty, we will apply the following trick.

Assume that the diagram indeed contains one ``harmful'' subgraph of the type
$\langle v'v'\rangle_{\rm 1-ir}$.
Since the $p$-dependence of the quantity $D^{(n)}(p)$ in (\ref{G3}) is known,
we can integrate the both sides of that equation over $\p$ with an arbitrary
``weight function'' $F(p^{2})$ (the only requirement is that the integral
be convergent) and then extract the amplitude $A^{(n)}$ from the identity
\begin{equation}
 g^{n+1} \mu^{2\eps(n+1)} \nu^2 A^{(n)}
= \frac {\int d{\bf p} D^{(n)}(p) F(p^{2})} {S_{d} \int_{0}^{\infty}
dp\, p^{1-2\eps(n+1)} F(p^{2})}.
\label{trick}
\end{equation}
In the denominator we have performed the angular integrations, which gives
the factor $S_{d}$ from (\ref{surface}). The numerator is represented by
the integral over $(n+1)$ momenta: $\p$ and $n$ independent momenta from the
original diagram. As is clear from the examples in figure~4, the number of
the lines $\langle vv\rangle _0$ in any $n$-loop diagram is equal to $(n+1)$.
Thus it is possible to perform the change of variables in the integral such
that the pure independent integration momenta will be assigned to all the
$\langle vv\rangle _0$ propagators. Now the powers of the integration
momenta $k^{-d}$ from the correlators (\ref{1.9}) and $k^{d}$ from
the Jacobians cancel each other, the
remaining dependence on $d$ comes only from the contractions of projectors
and vertices with further angular integrations. Thus we arrive at the
integrals of the type discussed in the previous section, and they can be
calculated for $d\to\infty$ in the same manner. In particular, all the
scalar products can be dropped, and the angular integrals in the numerator
of (\ref{trick}) give only the factor $S_{d}^{n+1}$; we are left with an
$(n+1)$-fold integral over the moduli.

Let us derive a more explicit expression for the resulting integrals
and check their independence of the weight function $F(p^{2})$. In the
new variables the former external momentum is certain linear combination
of the part of the new independent momenta:
\begin{equation}
\p= {\bf k}_1+{\bf k}_2+ \dots +{\bf k}_{l}
\label{kombi}
\end{equation}
with some $l \ge 2$. The remaining $m=n-l+1$ independent momenta will be
denoted as ${\bf q}_1,\dots {\bf q}_m$. Analysis of the diagrams shows that
$m$ is equal to the number of the 1-irreducible subgraphs of the form
$\langle v'vv\dots v \rangle_{\rm 1-ir}$, that is, subgraphs with one field
$v'$ and arbitrary number of the fields $v$. This is easily checked for the
examples given in figure~4: $n=l=m=1$ for the diagram {\it b} and $n=l=2$,
$m=1$ for the diagrams {\it d} and {\it e}.\footnote{This statement
does not apply to the graphs like {\it a} and {\it c} in figure~4,
which have no ``harmful'' $\langle v'v'\rangle_{\rm 1-ir}$
subgraph and involve no momenta of the type ${\bf k}$; the trick
is not needed for such diagrams.}
It is also clear that the momenta ${\bf q}_{\,i}$ here are exactly the same
as in the original variables, while ${\bf k}_i$ in (\ref{kombi}) are certain
linear combinations of $\p$ and the original momenta of the type ${\bf k}$.

Let us introduce the $l$-dimensional ``vector''
${\bf K} = \{k_1 ,k_2, \dots, k_l\}$ (where $k_i=|{\bf k}_{i}|\ge0$ for all
$i=1,\dots,l$), the corresponding unit vector
${\bf n} = {\bf K}/K$, and pass to the spherical coordinates:
\begin{equation}
\int_{0}^{\infty} dk_1 \dots \int_{0}^{\infty} dk_{l} =
\int_{0}^{\infty} dK K^{l-1} \int d{\bf n}.
\label{sfK}
\end{equation}
The integrations over the directions of the vector ${\bf n}$ are restricted
by the inequalities $n_{i}\ge 0$. For $d\to\infty$, the scalar products
can be dropped, so that
\[ p^{2} = ({\bf k}_1+{\bf k}_2+ \dots +{\bf k}_{l})^{2} \simeq
k_1^{2}+k_2^{2}+ \dots +k_{l}^{2} = K^{2}. \]
The remaining variables can be made dimensionless by the replacement
$q_{i}\to  w_{i}=q_{i}/K$ ($i=1, \dots, m=n-l+1$). In the new set
of variables ${\bf n}$, $w_{i}$ and $K$ only the latter is
dimensional. Thus the dependence of the integrand on $K$ is found
from the dimensional considerations as a known power, the integral over
$K$ factorizes out from the remaining integrals in the form
\begin{equation}
I_{n} =\int_{0}^{\infty}  dK K^{1-2\eps(n+1)} F(K^{2})
\label{sfA}
\end{equation}
and explicitly cancels the integral in the denominator of (\ref{trick}).
This confirms the independence of the result on the choice of the
weight function $F(K^{2})$.

Thus we are left with some integrals over the dimensionless variables
${\bf n}$ and $w_{i}$ of the general form
\begin{equation}
\int_{n_{i}\ge0} d{\bf n}\, \int dw_{1} \dots \int dw_{m}\
\psi ({\bf n}, w_{1}, \dots w_{m})
\label{psi}
\end{equation}
with a known dimensionless function $\psi$. They appear rather simple: all
the two-loop integrals are calculated exactly (that is, without the
expansion in the Laurent series in $\eps$) and are expressed in terms
of $\Gamma$ functions, trigonometric functions and simpler quantities.

From the calculation procedure described above one can conclude that
in the limit $d\to\infty$, the momenta of all the propagators
$\langle vv\rangle_{0}$ (but {\it not} those of the mixed propagators
$\langle vv'\rangle_{0}$!) behave as if they were orthogonal to each other,
including the case when one of such propagators corresponds to the
external momentum ${\bf p}$. For any given diagram it can be checked
{\it a posteriori\,} that the condition that all such momenta be orthogonal
defines the stationarity manifold in the integration region in the
steepest-descent calculation for $d\to\infty$. However, it would be
difficult to derive the orthogonality condition starting from the
stationarity equations, especially in the original variables, where some
of the propagators $\langle vv\rangle_{0}$ correspond to linear combination
of ``pure'' integration momenta. Thus the trick described above appears
extremely useful.

Let us illustrate the general scheme by the explicit calculation of the
diagram {\it d} from figure~4. For this diagram $n=l=2$ and $m=1$ and,
in agreement with the general statements made above, it involves two
momenta of the type ${\bf k}$ in (\ref{kombi}) and one momentum ${\bf q}$.
All these momenta are assigned to the lines $\langle vv\rangle_{0}$ as
shown in figure~5; the remaining lines correspond to their linear
combinations. The dotted line denotes the weight function $F(p^{2})$,
where the former external momentum is represented as
${\bf p}={\bf k}_{1}+{\bf k}_{2}$.

\begin{figure}
\begin{center}
\includegraphics[height=20ex]{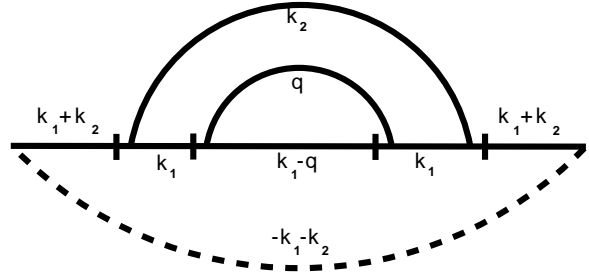}
\caption{The new choice of the independent integration momenta
in a two-loop diagram with the ``harmful'' subgraph.}
\end{center}
\end{figure}

The integrand in the numerator of the expression (\ref{trick}) has the form
\[ \left(-k_1^2 K^2\right)\times
\frac{3}{16\,K^4} \left\{ \frac{2}{k_1^2(k_1^2+q^2)}
+\frac{1}{(k_1^2+q^2)(K^2+q^2)}+\frac{1}{K^2(K^2+q^2)} \right\} \times \]
\[ \times\frac{1}{8(k_1k_2q)^{d-2+2\varepsilon}}, \quad
K^2 = k_1^2+k_2^2.  \]
Here the first factor comes from the vertices (\ref{vertexA}),
the second one -- from the time integrations and the third one -- from
the propagators $\langle vv \rangle_{0}$ in (\ref{linesV}).
After the angular integrations the integral in the numerator of
the expression (\ref{trick}) takes on the form
\begin{eqnarray}
J  &=& - \frac{S_d^3} {128\,(2\pi)^{2d}} \, \int_0^\infty dk_1
\int_0^\infty dk_2 \int_0^\infty dq\,
\frac{k_1^2}{K^2(k_1k_2q)^{2\varepsilon-1}} \times   \nonumber \\
&\times&
\left\{ \frac{2}{k_1^2(k_1^2+q^2)}+\frac{1}
{(k_1^2+q^2)(K^2+q^2)}+\frac{1}{K^2(K^2+q^2)}\right\} F(K^2).
\end{eqnarray}
Passing to the new variables $k_i=Kn_i$, $q=Kw $ gives
\[ J = \frac{S_d^3}{128\,(2\pi)^{2d}}\ \tilde JI _{0}, \]
where the integral
\[ I_{0} =\int_0^\infty dK K^{1-2\varepsilon}F(K^2) \]
from (\ref{sfA}) cancels out the denominator
in (\ref{trick}) and the remaining integral $\tilde J$ has the form
\begin{eqnarray}
\tilde J &=& - \int_{n_i\ge 0} d\n
\int_0^\infty dw\, \frac{n_1^2}{(n_1n_2w)^{2\eps-1}}
\left\{\frac{2}{n_1^2(n_1^2+w^2)}+ \right.   \nonumber \\
 &+& \left. \frac{1}{(n_1^2+w^2)
(1+w^2)}+\frac{1}{(1+w^2)} \right\}.
\label{aik}
\end{eqnarray}
In the spherical coordinates $n_{1} = \cos\theta$ and $n_{2} = \sin\theta$
the integral over the directions in (\ref{aik}) takes on the form
\[ \int_{n_i\ge 0} d\n=\int_0^{\pi/2}d\theta. \]
The remaining integrations over $w$ and $\theta$ are performed explicitly
and give the following final answer for the diagram in figure~5:
\begin{equation}
\tilde J= \Gamma(\varepsilon)\,\Gamma(-\varepsilon)\,\Gamma(1-\varepsilon)
\left\{\frac{(4\varepsilon-1)\,\Gamma(1-2\varepsilon)}
{2\,\Gamma(2-3\varepsilon)}+\frac{(2-\varepsilon)\,
\Gamma(1-\varepsilon)}{4\,\Gamma(2-2\varepsilon)}\right\}.
\label{Pes}
\end{equation}

The explicit result (\ref{Pes}) illustrates some general properties of the
diagrams of the pair correlator (\ref{G}). In general, the diagrams contain
poles at $\eps=0$ (manifestation of the UV divergences) which are cancelled
by the poles coming from the renormalization constants in the action
(\ref{Renact}) in every order of the renormalized perturbation theory, so
that the function $R$ in (\ref{dimG}) is finite at $\eps\to0$ in every order
of the expansion in $u$. In addition, the expression (\ref{Pes}) involves
the poles for some finite values of $\eps$, namely $\eps=1/2$, $1$, $3/2$
and $2$. All the other two-loop diagrams contain the poles at $\eps=1$ and
$2$ and some more diagrams contain the poles at $\eps=1/2$ and $3/2$. Such
singularities are manifestation of the IR divergences (that is, divergences
coming from the region of small integration momenta) that occur in the
diagrams if the calculation is performed in the ``massless'' ($m=0$) model.
When the IR cut-off $m>0$ in (\ref{1.9}) is restored, the diagrams become
convergent for all $\eps>0$, and the IR problem is reformulated as the
analysis of the singularities at $m\to0$. The problem of such singularities
in the stochastic NS equation with the stirring force (\ref{1.9}) was
discussed in a number of studies; see e.g. \cite{K}--\cite{Kim2}.

The one-loop diagrams only contain the first-order poles at $\eps=1$ and $2$.
As illustrated by the present two-loop calculation, in the higher-order
diagrams the order of such poles increases and they are getting closer and
closer to the origin $\eps=0$. Thus the problem of the IR singularities lies
beyond the scope of any given-order approximation of the RG, let alone the
ordinary perturbation theory. For the correct analysis of the small-$m$
behaviour, the standard RG techniques should be combined with additional
methods -- infrared perturbation theory or short-distance operator-product
expansion \cite{APS}--\cite{Kim2}. For small values of $\eps$, this approach
allows one to prove the existence of the limit $m\to0$, which justifies the
expansion in $\eps$ of the scaling amplitude $R(1,u_{*})$ in
(\ref{dimG}).\footnote{More physically, the IR singularities at $\eps=1/2$
and $1$ are related to the so-called sweeping effects and must disappear
in the sum of all the diagrams for the equal-time correlator, which is
a Galilean-invariant quantity; they do. The singularities at $\eps=3/2$
and $2$ do not disappear; they can be related to the fluctuations of the
energy flux. These interesting issues will be discussed elsewhere.}

The final result for the scaling function in (\ref{dimG}) has the form
\begin{equation}
R(1,u) = \frac{1}{2} + \frac{u}{16} \left\{ \frac{1}{2} + \eps
\left(1 - \frac{\pi^2}{6} \right)\right\} + \frac{u^{2}\pi^2}{384} + \dots,
\label{R1u}
\end{equation}
up to the terms of higher orders in $u \sim \eps$. Substituting the
fixed-point coordinate (\ref{ufix3}) gives
\begin{equation}
R(1,u_{*}) = 1/2 + \eps/12 + (5/36 - \pi^2 /108) \eps^{2} + O(\eps^{3}).
\label{EpsEx}
\end{equation}
This is the main result of the present section; it will be used below in
the calculation of the Kolmogorov constant and the skewness factor.
Although only the first two terms of the coordinate (\ref{ufix3}) are
involved in the derivation of (\ref{EpsEx}), the $O(\eps^{3})$ term will
be needed to calculate the three terms of the whole expression (\ref{dimG}),
which contains the product $u_{*}^{1/3}R(1,u_{*})$. In this sense, the
two-loop calculation of the pair correlator is consistent with the
three-loop calculation of the RG functions.

\section{Two-loop calculation of the Kolmogorov constant
and skewness factor} \label{sec:CK}

The Kolmogorov constant $C_{K}$ can be defined as the dimensionless
coefficient in the inertial-range asymptotic expression
$S_{2}(r)=C_{K}(\E r)^{2/3}$ for the second-order structure function,
predicted by the classical phenomenological Kolmogorov--Obukhov theory
\cite{Legacy,Monin}.\footnote{Now it is generally believed that the real
exponent deviates slightly from $2/3$ due to the phenomenon of
intermittency; see the discussion in \cite{Legacy} and references therein.
However, the situation is not absolutely clear: some researchers argued
that the existing experimental data are consistent with the ``$2/3$ law''
and the observed disagreement is due to the {\it corrections} (and not
{\it deviations}) vanishing in the limit of infinite Reynolds number;
see e.g. \cite{Lund}. Now we are interested in the large-$d$ limit, in
which intermittency is expected to get weaker or to disappear; so we will
accept the ``$2/3$ law,'' which is also internally consistent within the
calculation based on the $\eps$ expansion.}
Here $\E$ is the average energy dissipation rate (cf. eq. (\ref{1.3})) and
the $n$-th order (longitudinal, equal-time) structure function is defined as
\begin{equation}
S_{n} (r) = \big\langle [ v_{r} (t,{\bf x}+{\bf r})
- v_{r} (t,{\bf x})]^{n} \big\rangle, \qquad
v_{r} = (v_{i} r_{i})/r, \quad r= |{\bf r}|.
\label{struc}
\end{equation}
Using this definition, the function $S_{2}$ can be related to the
momentum-space pair correlation function $D(k)$ from (\ref{G}) as follows:
\begin{eqnarray}
S_{2}(r) = 2\int \frac{d{\bf k}}{(2\pi)^{d}}
\,D(k)\, \left[1-({\bf k}{\bf r})^{2}/(kr)^{2}\right]
\left\{1- \exp \left[{\rm i} ({\bf k}{\bf r})\right]\right\}.
\label{atas}
\end{eqnarray}
Alternatively, the Kolmogorov constant $C_{K}'$ can be introduced through
the phenomenological relation $E(k)=C_{K}' \E^{2/3} k^{-5/3}$, where the
energy spectrum $E(k)$ is related to the function (\ref{G})
as $E(k)= \bar S_{d}(d-1)k^{d-1}D(k)/2$. From the definitions one can
derive the following relation between these two constants:
\begin{equation}
C_{K} = \frac {3\cdot2^{1/3}\Gamma(2/3)\Gamma(d/2)}
{(d+2/3)\Gamma(d/2+1/3)}\, C_{K}' ,
\label{sviaz}
\end{equation}
cf. \cite{Monin} for $d=3$.
Using the exact relation $S_{3}(r)=-12\E r/d(d+2)$ that follows from the
energy balance equation (see e.g. \cite{Legacy,Monin} for $d=3$), the
constant $C_{K}$ can be related to the inertial-range skewness factor
${\cal S}$ as follows:
\begin{equation}
{\cal S} \equiv S_{3}/S_{2}^{3/2} = -[12/d(d+2)]\,C_{K}^{-3/2}.
\label{sviaz_2}
\end{equation}
All these relations refer to real physical quantities in the inertial
range, which in the stochastic model (\ref{1.1}), (\ref{1.2}) corresponds
to $\eps=2$ and $m=0$ in the random force correlator (\ref{1.9}).

Much work has been devoted to derivation of the Kolmogorov constant
within the RG approach; see e.g. \cite{APS,JETP,Lam,48,78,Giles} and
references therein. In order to obtain $C_{K}$, one usually combines
the RG expression (\ref{dimG}) for $D(k)$ or the analogous expression
for $S_{2}$ by some relation between the physical parameter $\E$ and
the amplitude $D_{0}$ in the random force correlator (\ref{1.9}).
In particular, in \cite{48} the first-order term of the $\eps$ expansion
for the pair correlator was combined with the so-called eddy-damped
quasinormal Markovian approximation for the energy transfer function,
taken directly at the physical value $\eps=2$. More elementary derivation,
based on the exact relation (\ref{1.3}) between $\E$ and the function
$d_{f}(k)$ from (\ref{1.9}) was given in \cite{JETP}; see also
\cite{UFN,turbo}. In spite of a reasonable agreement with the experiment,
such derivations are not satisfactory from the theoretical viewpoints.
Their common drawback is that any relation between $\E$ and $D_{0}$ is
unambiguous only in the limit $\eps\to2$ (see equation (\ref{2.74})),
so that the coefficients of the $\eps$ expansions for $C_{K}$ can be made
arbitrary; see e.g. the discussion in \cite{Lam} and Sec.~2.10 of
\cite{turbo}.

This ambiguity is a consequence of the fact that the notion itself of
the Kolmogorov constant has no unique extension to the nonphysical range
$0<\eps<2$. Furthermore, the experience from the RG theory of critical
behaviour suggests that well-defined $\eps$ expansions can be written for
universal quantities, such as critical exponents or normalized scaling
functions, which do not involve bare parameters. The constant $C_{K}$
extended to the range $0<\eps<2$ as in Refs. \cite{JETP,48,78} involves a
bare parameter, $D_{0}$, and hence is not universal in this sense.

To circumvent these difficulties, an alternative derivation was proposed
in \cite{APS} that does not involve any relation between $D_{0}$ and $\E$,
relates $C_{K}$ to a universal (in the sense of the theory of critical
behaviour) quantity, and leads (for $d=3$) to reasonable agreement with
experimental data. Below we adopt this derivation to our case, $d\to\infty$.

Consider the ratio
\begin{equation}
Q(\eps) = {\cal D}_{r} S_{2}(r) / |S_{3}(r)|^{2/3}=
{\cal D}_{r} S_{2}(r) / (-S_{3}(r))^{2/3}
\label{RR}
\end{equation}
with ${\cal D}_{r} =  r\partial/\partial r$. We shall see below that
it is universal and can be calculated in the form of a well-defined $\eps$
expansion. On the other hand, its value at $\eps=2$ determines the
Kolmogorov constant and the skewness factor through the exact relations
\begin{equation}
C_{K}= \left[3Q(2)/2\right]\left[12/d(d+2)\right]^{2/3},
\quad
{\cal S} = - \left[3Q(2)/2 \right]^{-3/2}
\label{trud}
\end{equation}
which follow from the definitions, relation (\ref{sviaz_2}) and the identity
${\cal D}_{r} r^{\delta} = \delta r^{\delta}$ for any $\delta$.

From the RG equations one derives the analogs of the representation
(\ref{dimG}) for the structure functions entering the ratio (\ref{RR}):
\begin{equation}
S_{3}(r) = D_{0} r^{-3\Delta_{v}} f_{3} (\eps), \quad
{\cal D}_{r} S_{2}(r) = D_{0}^{2/3} r^{-2\Delta_{v}}
f_{2}(\eps).
\label{RGS}
\end{equation}
Here $m=0$, $\mu r\gg 1$, $0<\eps\le 2$ and $\Delta_{v} = 1-2\eps/3$,
cf. (\ref{dimG}). The operation ${\cal D}_{r}$
introduced in (\ref{RR}) ``kills'' the constant contribution
$\langle v^{2} \rangle$ in $S_{2}$ which does not exist without an UV
cutoff for $\eps<3/2$ (see below); in $S_{3}$ such contribution is absent.

It follows from (\ref{RGS}) that the amplitude $D_{0}$ disappears from the
ratio $Q(\eps)=f_{2}/(-f_{3})^{2/3}$, and the latter can be calculated
in the form $Q(\eps)=\eps^{1/3} p(\eps)$, where $p(\eps)$ is a power series
in $\eps$. Our results for the fixed point from section~\ref{sec:Grin}
and for the pair correlator from section~\ref{sec:pair} allow us to find
the first three terms of $p(\eps)$. In this sense, one can speak about the
third-order approximation for the Kolmogorov constant (previous attempts
have mostly been confined with the first order, with the second-order
exception of Ref.~\cite{APS}).
To avoid possible misunderstanding, we stress again that we did not intend
to extend the definition of the physical quantities $C_{K}$ and ${\cal S}$
to the whole interval $0<\eps<2$ and to construct their $\eps$
expansions from the known expansion for $Q(\eps)$. Instead, the latter is
used to give the value of $Q(2)$, which, in its turn, determines $C_{K}$
and ${\cal S}$ through the relations (\ref{trud}) that make sense only
for the real value $\eps=2$.

Applying the operation ${\cal D}_{r}$ to the expression (\ref{atas}) gives:
\begin{eqnarray}
{\cal D}_{r} S_{2}(r) = 2 \int \frac{d{\bf k}}{(2\pi)^{d}}\, D(k)\,
\left[1-({\bf k}{\bf r})^{2}/(kr)^{2}\right]\,
({\bf k}{\bf r}) \,\sin ({\bf k}{\bf r}).
\label{atas1}
\end{eqnarray}
In order to obtain the inertial-range form of the function
${\cal D}_{r} S_{2}(r)$ it is sufficient to substitute the asymptotic
expression (\ref{dimG}) into (\ref{atas1}). A straightforward
calculation gives:
\begin{eqnarray}
{\cal D}_{r} S_{2}(r) = \frac {2(d-1) \Gamma(2-2\eps/3)}
{(4\pi)^{d/2}\Gamma(d/2+2\eps/3)}\, g_{*}^{1/3} \,R(1,u_{*}) \,
D_{0}^{2/3} (r/2)^{-2\Delta_{v}}
\label{atas4}
\end{eqnarray}
with the amplitude $R(1,u_{*})$ from (\ref{dimG}). It is important here
that the integral in (\ref{atas1}) with the function (\ref{dimG}) exists
for all $0<\eps<2$. It is the operation ${\cal D}_{r}$ that ensures its
convergence: the original integral (\ref{atas}) would be UV divergent
for $0<\eps<3/2$.

The needed terms of the $\eps$ expansion for $f_{3}$ can be obtained from
the direct perturbative calculation, similar to the calculation of the pair
correlator in section~\ref{sec:pair}, but it is more convenient to use
the exact expression
\begin{equation}
S_3(r)= -\frac{3(d-1) \Gamma(2-\eps)}
{(4\pi)^{d/2}\Gamma(d/2+\eps)} \, D_{0} (r/2)^{-3\Delta_{\varphi}}
\label{Exa}
\end{equation}
which follows from the energy balance equation and generalizes the well-known
Kolmogorov's ``4/3 law'' \cite{Legacy,Monin} to general $d$ and $\eps$.
Thus from Eqs. (\ref{atas4}) and (\ref{Exa}) for the ratio $Q(\eps)$ one
obtains
\begin{eqnarray}
Q(\eps)= \left[4(d-1) u_{*} /9 \right]^{1/3}\, A(\eps)\, R(1,u_{*}),
\label{atas2}
\end{eqnarray}
where we excluded $g_{*}$ in favor of $u_{*}= g_{*} \bar S_d$;
the coefficient $A(\eps)$ is given by
\begin{eqnarray}
A(\eps) = \frac {\Gamma(2-2\eps/3)\Gamma^{1/3}(d/2)
\Gamma^{2/3}(d/2+\eps)} {\Gamma(d/2+2\eps/3)\Gamma^{2/3}(2-\eps)}.
\label{atas3}
\end{eqnarray}
Using the Stirling formula for the $\Gamma$ functions one can show that
\[ \frac {\Gamma^{1/3}(d/2)\Gamma^{2/3}(d/2+\eps)}
{\Gamma(d/2+2\eps/3)} = 1+ O(1/d), \]
so that in the limit $d\to\infty$ one obtains
\begin{eqnarray}
A(\eps) = \frac {\Gamma(2-2\eps/3)} {\Gamma^{2/3}(2-\eps)}=
1+ \frac{1}{9} \left\{ 1-\pi^2/6 \right\} \eps^{2} +O(\eps^{3}).
\label{atas444}
\end{eqnarray}
Substituting the expressions (\ref{ufix3}), (\ref{EpsEx})
and (\ref{atas444}) into (\ref{atas2}) gives
\begin{equation}
Q(\eps) = \frac{(4\eps d)^{1/3}}{3} \left\{ 1 + \frac{\eps}{18} +
\left(\frac{49}{162}-\frac{\pi^2}{27} \right) \eps^{2} +O(\eps^{3})
\right\}.
\label{Qe}
\end{equation}
Now the Kolmogorov constant $C_{K}$ and the skewness factor ${\cal S}$
are obtained from the expressions (\ref{trud}), with the replacement
$d(d+2) \to d^{2}$ in the first one. Thus in the leading order of the
large-$d$ asymptotic behaviour one obtains $C_{K}\propto 1/d$ and
${\cal S} \propto 1/d^{1/2}$. From the relation (\ref{sviaz}) it then
follows $C_{K}'\propto d^{1/3}$, in agreement with the earlier results
derived within the direct interaction approximation \cite{FFR} and within
the RG approach \cite{UFN,turbo}.

Although these results refer to the large-$d$ limit, one can try to use them
as some approximation to the real three-dimensional case. Substituting the
third-order result (\ref{Qe}) into the second relation from (\ref{trud}) and
setting $\eps=2$ and $d=3$ gives ${\cal S} \approx -0.73$, while the
experimental value recommended in \cite{Monin} is ${\cal S} \approx -0.28$.
Of course, one should not have expected a better agreement. Surprisingly
enough, the situation appears much better for the Kolmogorov constant.
Substituting the first, second and third approximations from (\ref{Qe})
into the first relation from (\ref{trud}), one obtains for $\eps=2$ and $d=3$
the following results:
\[ C_{K}^{(1)}\approx 1.75, \quad C_{K}^{(2)} \approx 1.94, \quad
C_{K}^{(3)}\approx 1.50, \]
all of them in a reasonable agreement with the experimental estimate
$C_{K} \approx 1.9$ recommended in \cite{Monin}. One can speculate that
the skewness factor which is related to the odd-order function $S_{3}$
and is a measure of anisotropy in the distribution of the velocity field
is more sensitive to the spatial dimension than the Kolmogorov constant,
related to the even-order function $S_{2}$.

\section{Conclusion} \label{sec:Con}

We have accomplished the complete three-loop calculation
of the main ingredients of the RG analysis of the $d$-dimensional stochastic
Navier-Stokes equation (\ref{1.1}), (\ref{1.2}) in the limit $d\to\infty$:
the renormalization constant $Z_{\nu}$, the $\beta$ function, the coordinate
of the fixed point $u_{*}$ and the ultraviolet (UV) correction exponent
$\omega$; these results are summarized in equations
(\ref{Znu3})--(\ref{omega3}). We have also calculated in the large-$d$
limit the correlation function of the velocity in the third order
of the $\eps$ expansion (two-loop approximation).

These results allowed us to derive the third-order answers for two
interesting physical quantities: the Kolmogorov constant
$C_{K}$ in the spectrum of turbulent energy and the inertial-range skewness
factor ${\cal S}$. The former is in a reasonable agreement with the
existing experimental data for the real three-dimensional turbulence.

The successful analytic calculation in the third order appeared feasible
due to drastic simplifications that we found in the large-$d$ limit: in
particular, most of the Feynman diagrams for the Green function in that
limit vanish, and the remaining ones are reduced to relatively simple
(and analytically calculable) integrals. Some more efforts and tricks were
required to derive the third-order results for the pair correlation function
of the velocity.

Although we did not succeed in finding the exact solution for $d=\infty$,
the simplifications that occur in the calculations in that limit and the
simple form of the obtained results suggest that this is not an
impossible task. Based on the three-loop answers, we proposed some
hypothetical expressions beyond the $\eps$ expansion which show what,
in principle, such exact results may look like.

We believe that the present results, as well as the calculational
techniques developed in their derivation, will be useful in the
further attempts of the construction of the systematical $1/d$
expansion for the fully developed turbulence.

\section*{Acknowledgments}
The authors are thankful to Michal Hnatich, Juha Honkonen,
Paolo Muratore Ginanneschi and Mikhail Nalimov for discussions.
We are especially indebted to late Professor Alexander Nikolaevich Vasiliev
for his constant interest to our research and his numerous and valuable
discussions and suggestions.
The work was supported in part by the Russian Foundation for Fundamental
Research (grant No~08-02-00125a), the Russian National Program
(grant No~2.1.1.1112) and the program ``Russian Scientific Schools''
(grant No~5538.2006.2). N~V~A thanks the Department of
Mathematics in the University of Helsinki for their kind
hospitality during his visits, financed by the project ``Extended
Dynamical Systems.''

\end{document}